\documentclass[conference]{IEEEtran}
\usepackage[available, functional]{ndssbadges}
\usepackage{lineno}
\usepackage[table]{xcolor}
\usepackage{subcaption}
\usepackage{hyperref}
\usepackage{calc}

\usepackage{amsmath,amssymb}
\usepackage{algorithm,algpseudocode}
\usepackage{multirow}
\usepackage{amsfonts}
\usepackage{xcolor}
\usepackage{amsmath}
\usepackage[compatibility=false]{caption}
\usepackage{booktabs}
\usepackage{enumitem}
\usepackage{listings}
\usepackage{cite}
\hypersetup{
    colorlinks=true,
    linkcolor=blue,
    urlcolor=blue,
    citecolor=blue
}

\definecolor{codebg}{RGB}{248,248,248}
\definecolor{codeframe}{RGB}{200,200,200}
\usepackage{tcolorbox}
\newtcolorbox{promptbox}{
  colback=white,
  colframe=black,
  boxrule=0.5mm,
  arc=2mm,
  top=2mm,
  bottom=2mm,
  left=2mm,
  right=2mm,
  width=\columnwidth
}

\begin{document}
%
\title{In-Context Probing for Membership \\ Inference in Fine-Tuned Language Models}


\author{
\IEEEauthorblockN{Zhexi Lu\textsuperscript{1},
Hongliang Chi\textsuperscript{1},
Nathalie Baracaldo\textsuperscript{2},
Swanand Ravindra Kadhe\textsuperscript{2},
Yuseok Jeon\textsuperscript{3},
Lei Yu\textsuperscript{1}}
\IEEEauthorblockA{\textsuperscript{1}Rensselaer Polytechnic Institute, Troy, NY, USA\\
Email: \{luz17, chih3, yul9\}@rpi.edu}
\IEEEauthorblockA{\textsuperscript{2}IBM Research, San Jose, CA, USA\\
Email: \{baracald, swanand.kadhe\}@ibm.com}
\IEEEauthorblockA{\textsuperscript{3}Korea University, Seoul, South Korea\\
Email: ys\_jeon@korea.ac.kr}
}


%


\IEEEoverridecommandlockouts
\makeatletter\def\@IEEEpubidpullup{6.5\baselineskip}\makeatother
\IEEEpubid{\parbox{\columnwidth}{
		Network and Distributed System Security (NDSS) Symposium 2026\\
		23 - 27 February 2026 , San Diego, CA, USA\\
		ISBN 979-8-9919276-8-0\\  
		https://dx.doi.org/10.14722/ndss.2026.240892\\
		www.ndss-symposium.org
}
\hspace{\columnsep}\makebox[\columnwidth]{}}

\maketitle
\begin{abstract}
Membership inference attacks (MIAs) pose a critical privacy threat to fine-tuned large language models (LLMs), especially when models are adapted to domain-specific tasks using sensitive data. While prior black-box MIA techniques rely on confidence scores or token likelihoods, these signals are often entangled with a sample’s intrinsic properties—such as content difficulty or rarity—leading to poor generalization and low signal-to-noise ratios. In this paper, we propose ICP-MIA, a novel MIA framework grounded in the theory of training dynamics, particularly the phenomenon of diminishing returns during optimization. We introduce the Optimization Gap as a fundamental signal of membership: at convergence, member samples exhibit minimal remaining loss-reduction potential, while non-members retain significant potential for further optimization. To estimate this gap in a black-box setting, we propose In-Context Probing (ICP)—a training-free method that simulates fine-tuning-like behavior via strategically constructed input contexts. We propose two probing strategies: reference-data-based (using semantically similar public samples) and self-perturbation (via masking or generation). Experiments on three tasks and multiple LLMs show that ICP-MIA significantly outperforms prior black-box MIAs, particularly at low false positive rates. 
We further analyze how reference data alignment, model type, PEFT configurations, and training schedules affect attack effectiveness. Our findings establish ICP-MIA as a practical and theoretically grounded framework for auditing privacy risks in deployed LLMs.
\end{abstract}

%
\IEEEpeerreviewmaketitle

\section{Introduction}
Large language models (LLMs) have rapidly advanced in their generalization capabilities, enabling deployment across a wide range of real-world tasks. In privacy-sensitive domains such as healthcare, law, and finance, public open-source base models (e.g., LLaMA~\cite{touvron2023llama2openfoundation}) are routinely fine-tuned on small, domain-specific proprietary datasets to improve task performance~\cite{min2023recent,savage2025fine}. However, this practice raises serious privacy concerns. A growing body of research has shown that LLMs are vulnerable to privacy attacks at various stages of the model pipeline—including pre-training, distillation, fine-tuning, and inference—via techniques such as data extraction~\cite{carlini2021extracting} and membership inference~\cite{shidetecting, kim2025EM-MIA}. Among these stages, fine-tuning is particularly susceptible to privacy leaks, due to the typically limited size and sensitive nature of the fine-tuning datasets~\cite{yu2021differentially}.

Membership inference attacks (MIAs) aim to determine whether a particular data sample was part of a model’s training dataset, thereby potentially revealing sensitive or personally identifiable information about individuals~\cite{shokri2017membership}. MIAs have been extensively studied across a range of machine learning domains to identify and characterize privacy risks. These include generative adversarial networks (GANs)~\cite{chen2020gan}, explainable machine learning models~\cite{liu2024please}, and diffusion models~\cite{matsumoto2023membership}. In addition to vulnerability assessment, MIAs have also been employed to evaluate the efficacy of privacy-preserving mechanisms~\cite{wang2023lds,rezaei2023accuracy}, benchmark machine unlearning methods~\cite{kurmanji2023unboundedmachineunlearning}, and enable privacy auditing in deployed systems~\cite{mireshghallah2022quantifying,kazmi2024panoramia,wang2025membership}.

Recently, a growing body of work~\cite{mireshghallah2022empirical,duan2024membership} has adapted MIAs to assess the privacy risks of LLMs. These efforts build on classical MIA techniques but tailor them to the unique properties of LLMs. Broadly, existing MIA methods for LLMs can be categorized into two classes: reference-based attacks, which rely on an auxiliary/reference dataset (typically drawn from a distribution similar to the model's training data) to train one or more reference models, and reference-free attacks, which avoid this requirement.
Reference-based attacks, such as those by Mireshghallah et al.~\cite{mireshghallah2022empirical,mireshghallah2022quantifying}, extend the Likelihood Ratio Attack (LiRA) framework~\cite{carlini2022membership} to LLMs and masked language models. These attacks require training an ensemble of shadow models and comparing the target model’s negative log-likelihood (NLL) on a given sample against the distribution of NLLs from these shadow models to determine the membership. However, such approaches assume access to auxiliary data from the same distribution as the target model’s training set, an assumption that rarely holds in real-world scenarios, especially when fine-tuning involves private or proprietary datasets.
In contrast, reference-free attacks such as Min-K\%~\cite{shidetecting} and Min-K\%++~\cite{zhang2024min} detect memorized samples by identifying low-rank (outlier) tokens in the model’s output, which are indicative of overfitting. ReCaLL~\cite{xie-etal-2024-recall} demonstrates that adding a context prefix to the input can differentially affect the model’s predictions for memorized versus non-memorized samples. These attacks do not require access to reference data, making them significantly more practical for evaluating fine-tuned LLMs in real-world settings.

Despite recent advances in reference-free MIAs against LLMs, existing approaches largely lack a principled grounding, especially regarding training dynamics and memorization behavior. A key limitation is that these attacks rely heavily on raw confidence or loss values, which are strongly influenced by the intrinsic properties of each sample—such as its difficulty. As a result, current methods struggle to explain why certain tokens or perturbations reveal membership signals, which restricts their generality and robustness, and leads to degraded effectiveness.

\begin{figure}
    \centering
    \includegraphics[width=0.85\linewidth]{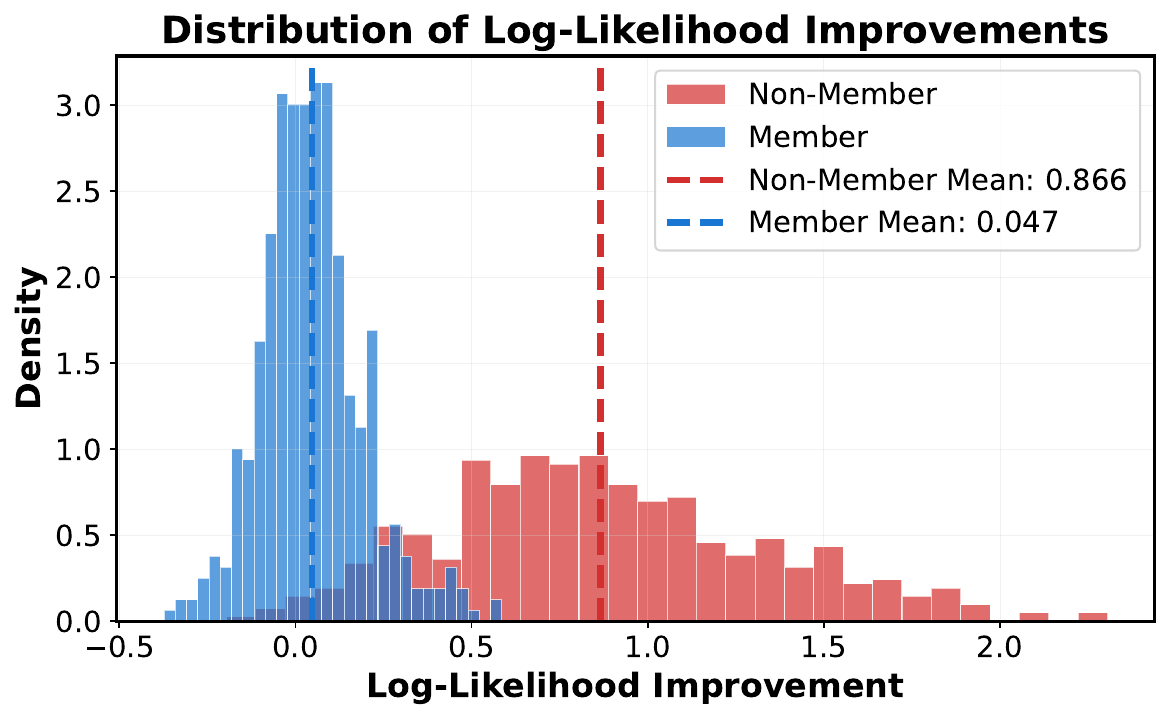}
    \caption{Log-likelihood improvement distribution on the HealthcareMagic dataset. Member samples (blue) show minimal gains from in-context probing, while non-members (red) exhibit larger, more variable improvements, revealing the optimization gap that underlies our attack.
    }
    \label{fig:Log_likelihood_Improvement}
    \vspace{-5pt}
\end{figure}

In this paper, we propose ICP-MIA, a novel membership inference framework grounded in the training dynamics of large language models. Prior work has shown that training samples typically experience rapid loss reduction during the early stages of fine-tuning, followed by diminishing returns as training progresses~\cite{mosbachstability}. This empirical pattern aligns with broader theoretical insights showing that models adapt quickly to seen data but converge more slowly with continued optimization~\cite{domhan2015speeding, viering2022shape}. Building on this observation, we identify the ``optimization gap''—the remaining loss-reduction potential of a sample at the end of fine-tuning—as a principled signal of membership. Our approach exacts this signal by using in-context probing to emulate a fine-tuning step at inference time. By observing how much the model’s confidence on a sample improves under these probing contexts, we obtain a practical black-box estimate of its optimization gap, enabling reliable separation between member and non-member samples. As illustrated in Figure~\ref{fig:Log_likelihood_Improvement}, using such a signal (i.e., log-likelihood improvement) significantly enhances the separation between member and non-member distributions.
To further enhance the robustness and practicality of ICP-MIA, we introduce two complementary strategies for constructing probes: a reference-data–based method that selects semantically aligned contexts from external datasets, and a reference-data–free method that generates self-perturbed probes using only the target sample. Together, these strategies strengthen ICP-MIA's effectiveness across diverse data distributions and threat models.

In summary, our main contributions are:
\begin{itemize}
\item \textbf{A Novel Framework for MIA.} We are the first to propose and formalize the \textbf{Optimization Gap}—the disparity in remaining loss-reduction potential between member and non-member samples—as a fundamental signal for membership inference. Our code can be found at \url{https://github.com/RPI-DSPlab/ICP-MIA}.
\item \textbf{A practical black-box method to estimate the Optimization Gap.} We introduce In-Context Probing (ICP), a training-free mechanism that simulates fine-tuning behavior at inference time. ICP-MIA includes two complementary strategies: a reference-data-based method that selects semantically aligned contexts from external data, and a reference-free self-perturbation method that eliminates the need for auxiliary datasets.
\item \textbf{State-of-the-Art Performance in Realistic Scenarios.}  We evaluate ICP-MIA on multiple LLMs and datasets, demonstrating consistent improvements over prior black-box attacks. On the HealthcareMagic dataset, ICP-MIA achieves an AUC of 0.942, surpassing ReCaLL (0.847) and Min-K\% (0.837). On CNN-DM, our method reaches a TPR@1\%FPR of 0.518, more than 2.6X higher than reference-free methods like ReCaLL (0.195), highlighting its effectiveness in high-precision scenarios.\end{itemize}

\section{Related Work}
\subsection{Membership Inference Attack against LLMs}
In this paper, we focus on MIAs in the black-box setting, where an attacker can only query the model and observe its output logits. The existing black-box MIAs can be categorized into two types:

\paragraph{Reference-based Attack} The classical Likelihood Ratio Attack (LiRA)~\cite{carlini2022membership} introduces statistical calibration to eliminate interference from the sample difficulty. It trains an ensemble of shadow models to estimate the NLL distributions of member and non-member samples, and compute the ratio between them as the membership score. Although effective, LiRA is computationally expensive due to the need for LLM shadow-model training.
A lighter approach trains a single reference model on public auxiliary data to approximate non-member behavior. This approach, however, depends critically on how closely the auxiliary data distribution matches the private fine-tuning distribution—an assumption that often fails in real-world settings. To address this mismatch, SPV-MIA~\cite{fu2024practicalmembershipinferenceattacks} uses self-prompt calibration, generating synthetic data from the target model to train a reference model and comparing the sample’s probabilistic variations under the two models to infer membership. While effective, this strategy may be infeasible under restrictive query budgets. DF-MIA~\cite{huang2025df} proposes a two-stage framework for fine-tuned LLMs that use test samples in the evaluation as a reference dataset and fuses reference-free and reference-based attacks. It first scores samples with a reference-free attack to build an augmented dataset that strengthens non-member cues, then trains a reference model on this data and applies a loss calibration attack. However, relying on test samples as the reference dataset is a strong assumption and may introduce bias into the evaluation.

\paragraph{Reference-Free Attack} 
The most fundamental of reference-free approach is the Loss Attack, which uses a sample's negative log-likelihood (NLL) as a membership score. It is based on the principle that models are generally more confident on training data, meaning members should have a lower NLL on average~\cite{yeom2018privacy}. However, its reliability suffers from the intrinsic properties of samples; for instance, an easy non-member can have a lower loss than a difficult member. To mitigate this problem,  Min‑K\%~\cite{shidetecting} focuses on the "surprising" tokens by averaging the negative log-likelihood over the lowest-probability k\% tokens. Min‑K\%++~\cite{zhang2024min} further improves this signal by standardizing each token’s log-likelihood relative to the model’s conditional distribution, and then aggregating the most extreme standardized values. However, DC-PDD~\cite{zhang2024pretraining} claims that non-member texts composed of common high-frequency tokens may be misclassified as training data. Neighborhood Attack~\cite{mattern2023neighborhood} compares a sample’s score to those of its perturbed neighbors, based on the observation that neighbors of members tend to show larger confidence drops. 

ReCaLL~\cite{xie-etal-2024-recall} showed that prefixing target samples with non-member context causes a greater reduction in log-likelihood for member data than for non-member data, creating a distinctive asymmetric signal for membership inference. 
CON-RECALL~\cite{wang2025conrecall} amplifies this effect by contrasting member-style and non-member-style prefixes, while EM-MIA~\cite{kim2025EM-MIA} improves robustness by optimizing prefix quality.
While these approaches outperform raw loss attacks, they are largely empirical and rely on LL shifts induced by non-member prefixes. 
In contrast, our ICP-MIA approach also uses prefixes but treats them as in-context \emph{demonstrations} that simulate an additional fine-tuning step—fundamentally different from ReCaLL. This grounding in the model’s residual learning potential yields a more principled, robust, and interpretable membership signal.

MIAs have also been extended to diffusion models and multimodal models. Unlike LLMs where membership can be inferred directly from token-level probabilities, vision–language models (VLLMs) lack discrete ground-truth tokens for image inputs. To address this, Li et al.\cite{li2024membership} introduced a cross-modal pipeline that uses the model’s generated text descriptions as proxies, and proposed the MaxRényi-K\% metric to capture the model’s elevated confidence on key tokens when describing memorized images. Pang et al.\cite{pangblack} studied MIAs against fine-tuned diffusion models in black-box settings where log probabilities are unavailable. Their approach measures membership by evaluating the similarity between reconstructed images and their ground-truth counterparts, demonstrating that members generally yield substantially higher reconstruction similarity.

\subsection{In-Context Learning}
LLMs demonstrate a powerful capacity for In-Context Learning (ICL), where they adapt to tasks using prompted examples without explicit parameter updates. Recent studies increasingly characterize ICL as a form of implicit fine-tuning, where models leverage context to perform learning-like computations. Dai et al.~\cite{dai2023WhycanGPT} frames LLMs as “meta-optimizers,” where the Transformer attention mechanism computes “meta-gradients” from context examples, effectively creating a temporary, task-specific model. This view is further supported by Akyürek et al.~\cite{akyureklearning}, who demonstrate that Transformers can implicitly simulate learning algorithms, such as gradient descent and ridge regression, directly within their forward pass. Chen et al.~\cite{chenbypassing} extend this understanding by showing that looping Transformers can efficiently execute multi-step gradient descent. These studies establish that ICL can be interpreted as an implicit optimization process, enabling an LLM to dynamically adapt to tasks using context alone. 

Designing effective ICL prompts, however, is non-trivial, so researchers have explored in-context probing (ICP) as an alternative technique. Unlike ICL, which uses context to adapt the model’s predictions, ICP directly measures the influence of a given context on a target sample’s log-likelihood, offering a more precise analytical method. Zhuo et al.~\cite{zhoudetail} introduce an influence-function-based method to attribute model predictions to specific in-context examples. Amini et al.~\cite{amini2023context} further validate its effectiveness in measuring the impact of specific training-like exposures. Building on these insights, Jiao et al.~\cite{jiao2025feasibilityincontextprobingdata} demonstrate that ICP can approximate gradient-based influence functions without accessing model gradients, particularly when the probe context shares task or content similarity with the target.

\subsection{Fine‑Tuning Methods for Large Language Models}
Fine-tuning LLMs adapts pretrained models to downstream tasks by updating some or all of the model parameters. While this enables task-specific adaptation, the choice of fine-tuning strategy significantly impacts efficiency, scalability, and performance. Full fine-tuning (updating all model weights) offers maximum flexibility and strong performance gains. However, it is often impractical for large-scale models due to high computational and memory costs.

To address these challenges, a family of \emph{Parameter-Efficient Fine-Tuning} (PEFT) techniques has emerged. These methods selectively update a small subset of parameters or introduce auxiliary modules, significantly reducing training overhead while preserving most of the pretrained knowledge. Among PEFT methods, \emph{adapter-based fine-tuning} inserts lightweight trainable modules (adapters) into each transformer layer while freezing the original model weights. These adapters encode task-specific transformations, allowing a modular and efficient way to specialize models.
Another widely adopted PEFT method is \emph{Low-Rank Adaptation} (LoRA)~\cite{hu2021loralowrankadaptationlarge,liu2024dora,hayou2024lora+}. LoRA injects low-rank matrices into existing weight tensors, enabling efficient updates with minimal parameter growth. It achieves strong performance with reduced memory usage and no added inference latency, making it ideal for resource-constrained environments.

Another family of PEFT methods focuses on modifying inputs rather than model weights. Prompt-Tuning~\cite{lester2021powerscaleparameterefficientprompt} optimizes continuous prompt embeddings, while Prefix-Tuning~\cite{li2021prefixtuningoptimizingcontinuousprompts} and P-Tuning~\cite{liu2022ptuningv2prompttuning} extend this idea by injecting trainable prefix vectors at deeper model layers. These techniques maintain frozen backbone weights and adapt only a small embedding space, offering high flexibility with low memory cost.

Finally, quantization-based PEFT combines low-precision weights with parameter-efficient updates. QLoRA~\cite{dettmers2023qloraefficientfinetuningquantized} uses 4-bit quantization together with LoRA adapters, enabling fine-tuning of very large models on commodity hardware while maintaining near–full-precision accuracy.

\section{Problem Statement}

\subsection{LLM Supervised Fine-tuning}
This paper focuses on supervised fine-tuning (SFT), the standard approach for adapting a pre-trained LLM (e.g., LLaMA, GPT) to downstream tasks using labeled training examples.
Let the pre-trained model be denoted as $\mathcal{M}$ with parameters $\theta_{\text{pre}}$. The objective of fine-tuning is to obtain updated parameters $\theta_{\text{ft}}$ that minimize the negative log-likelihood (NLL) loss over a dataset $D_{\text{train}}= \{s_1, s_2, \dots, s_N\}$.

For generality, we denote each training example as a pair $s = (x, y)$, where $x$ is the task input (e.g., a prompt, document, context, or question) and $y$ is the expected output (e.g., a response, summary, or code). This covers a wide range of SFT scenarios, including instruction tuning, summarization, question answering, and domain-specific modeling. In instruction tuning, x represents a natural-language instruction and y the target response. For tasks such as summarization, QA, or dialogue generation, x may be a document or query, and y the corresponding output (e.g., a summary or answer).

For each sample $s_i=(x_i, y_i)$, fine-tuning minimizes the conditional NLL of generating the target output $y_i$ given the
input $x_i$:
\begin{equation}
\mathcal{L}(s_i, \theta) = - \sum_{t=1}^{|y_i|} \log p_{\theta}(y_{i,t} \mid x_i, y_{i,<t})
\end{equation}
where $p_{\theta}(y_{i,t} \mid x_i, y_{i,<t})$ is the model’s predicted probability of the $t$-th token in $y_i$, conditioned on the full input prompt $x_i$ and previous tokens $y_{i,<t}$. It is a common practice in SFT to compute the loss exclusively on the response tokens ($y_i$), while the input tokens ($x_i$) are masked out during the loss calculation~\cite{huerta-enochian-ko-2024-instruction}.
The goal of fine-tuning is to learn parameters $\theta_{\text{ft}}$ that minimize the total loss across all samples in $D_{\text{train}}$. Starting from the pre-trained weights $\theta_{\text{pre}}$, an optimizer such as SGD or Adam updates the parameters to solve:
\begin{equation}
\begin{split}
\theta_{\text{ft}} &= \arg \min_{\theta} \sum_{(x_i, y_i) \in D_{\text{train}}} \mathcal{L}((x_i, y_i), \theta) \\
&= \arg \min_{\theta} \sum_{(x_i, y_i) \in D_{\text{train}}} \left( - \sum_{t=1}^{|y_i|} \log p_{\theta}(y_{i,t} | x_i, y_{i, <t}) \right)
\end{split}
\end{equation}

The resulting model $\mathcal{M}_{\text{ft}}$($\theta_{\text{ft}}$) is then deployed for downstream applications. Our goal is to assess its susceptibility to membership inference based on how it responds to previously seen versus unseen prompt–response pairs.

\subsection{Threat Model and MIA Formulation}
\label{ssec:problem}
Given the target model $\mathcal{M}_{\text{ft}}$ obtained by supervised fine-tuning of a pre-trained base model $\mathcal{M}_{\text{pre}}$ on a private dataset $D_{\text{train}}$, 
we consider a black-box threat model where the adversary aims to determine if a sample $s \in D_{\text{train}}$ or  $s \notin D_{\text{train}}$, but while only having query access to 
$\mathcal{M}_{\text{ft}}$, and no ability to inspect or modify its internal parameters. 

Following state-of-the-art MIAs against LLMs~\cite{shidetecting,mattern2023neighborhood,xie-etal-2024-recall},
we assume that  the model API exposes per-token log-probabilities $\{\log p_{\theta}(y_t | x, y_{<t})\}_{t=1}^{|y|}$ for any input–output pair $(x, y)$.  This reflects realistic deployments, as many production APIs (e.g., Gemini, Grok) and self-hosted open-source models served via frameworks such as vLLM~\cite{kwon2023efficient} or Hugging Face Transformers~\cite{wolf2020transformers} provide this level of access. We further assume that samples in $D_{\text{train}}$ are not contained in the pre-training corpus of the base model $\mathcal{M}_{\text{pre}}$. This is consistent with common practice in real-world deployments where organizations fine-tune open-source base models (e.g., LLaMA) on proprietary datasets, such as medical, legal, or enterprise records, that are distinct from the publicly sourced data used during pre-training.
We discuss weaker assumptions, including label-only access (only predicted tokens are returned, without probabilities) and the case in which target samples appear in the pre-training corpus in Section~\ref{sec:discussion}.

\noindent\textbf{MIA Game.} Following the framework in prior MIA literature
~\cite{yeom2018privacy, carlini2022membership}, we can formalize MIA as a security game $\mathcal{G}_{\text{MIA}}$ between a challenger $\mathcal{C}$ and an adversary $\mathcal{A}$ as follows:
\begin{itemize}
    \item The challenger $\mathcal{C}$ samples a dataset \( D_{\text{train}}  {\leftarrow} \Omega^n \) from distribution $\Omega$. Then $\mathcal{C}$ obtains a fine-tuned model $\mathcal{M}_{\text{ft}}$ by fine-tuning a base model on $D_{\text{train}}$.
    \item The challenger draws a secret $b {\leftarrow} \left\{0, 1\right\}$. If $b = 1$, the challenger samples a point $s$ uniformly from $D_{\text{train}}$. If $b = 0$, the challenger samples $s$ from $D_{\text{test}}$, a hold-out set disjoint from $D_{\text{train}}$.
    \item Given $s$, the adversary $\mathcal{A}$ queries $\mathcal{M}_{\text{ft}}$ through API access to compute a membership score $\text{Score}(s, \mathcal{M}_{\text{ft}})$, and makes a binary prediction by thresholding the score:
\begin{equation}
\mathcal{A}(s, \mathcal{M}_{\text{ft}}) = 
\begin{cases} 
1 & \text{if } \text{Score}(s^*, \mathcal{M}_{\text{ft}}) > \tau \\
0 & \text{otherwise}
\end{cases}
\end{equation}
    \item The adversary wins the game if $\mathcal{A}(s, \mathcal{M}_{\text{ft}}) = b$.
\end{itemize}

Typical metrics for evaluating MIA effectiveness include \emph{AUC (Area Under the ROC Curve)}, which measures the attack’s overall discriminative power across all thresholds, and \emph{TPR@low FPR}, which reports the true positive rate at a fixed low false positive rate (e.g., 1\%) and reflects performance in high-precision settings where false positives are costly.

\section{Bridging Training Dynamics and Membership Signals via In-Context Probing}
\label{sec:Bridging}
In this section, we establish a theoretical and empirical foundation connecting the dynamics of neural network training with membership inference. Specifically, we introduce and validate the \textbf{Optimization Gap} as a fundamental signal of membership and demonstrate how \textbf{In-Context Probing (ICP)} can effectively approximate this gap in a black-box setting.

\subsection{The Optimization Gap: A Fundamental Membership Signal}
Existing MIAs typically rely on assumptions about post-training model behavior, using metrics such as model confidence or likelihood scores. However, these signals are heavily influenced by each sample’s inherent difficulty or uniqueness, making it hard to tell whether a high or low score reflects memorization or simply the nature of the input.
In this paper, we propose examining fundamental training dynamics, specifically the well-known phenomenon of diminishing returns, to overcome these limitations. Empirically, neural networks make fast progress early in training: the loss drops quickly at first and then slows down, eventually flattening out. Prior work models this pattern using a power-law decay~\cite{DBLP:conf/icml/BordelonAP24,hoffmann2022training,kaplan2020scaling}:
\[
\mathcal{L}(t) \approx C_t\,t^{-\alpha_t} + \mathcal{L}_{\infty},
\]
Here, $\mathcal{L}(t)$ denotes the loss at training step $t$, $\alpha_t$ controls how fast it falls, and $\mathcal{L}_{\infty}$ is the lowest loss the model approaches. This form explicitly captures the ``diminishing returns'' of training—large improvements early on and much smaller improvements later.

Motivated by this training behavior, we define the \emph{Optimization Gap} of a sample $(x,y)$ as the amount of loss the model could still reduce if it continued training on that sample, i.e.,
$$
\mathcal{L}(\theta^*; x,y) - \mathcal{L}(\theta; x,y)
$$
where the first term is the loss under the model $\theta^*$, and the second term is the loss under the model $\theta$ obtained by applying one additional optimization step to $\theta^*$.
At convergence, member samples have already been optimized during fine-tuning and therefore exhibit little or no remaining improvement, resulting in a small optimization gap. In contrast, non-member samples still have room for loss reduction, producing a noticeably larger gap.

Evidence from prior work further supports this hypothesis in the context of LLM fine-tuning. Komatsuzaki et al.\cite{komatsuzaki2019epochneed} show that most improvements occur in the first epoch of fine-tuning, with subsequent epochs providing sharply diminishing benefits. Similar observations are reported in Devlin et al.\cite{devlin2019bertpretrainingdeepbidirectional} and in later studies~\cite{dodge2020finetuningpretrainedlanguagemodels,xue2023repeatrepeatinsightsscaling}, which collectively indicate that 1–3 epochs are typically sufficient for effective LLM fine-tuning, and further training may even harm performance by overfitting. These findings reinforce our use of the optimization gap as a membership signal: \emph{the fine-tuning process inherently creates a sharp separation between previously seen and unseen samples in terms of their residual optimization potential.}

\begin{figure}
    \centering
        \centering
        \includegraphics[width=0.8\linewidth, height=5cm]{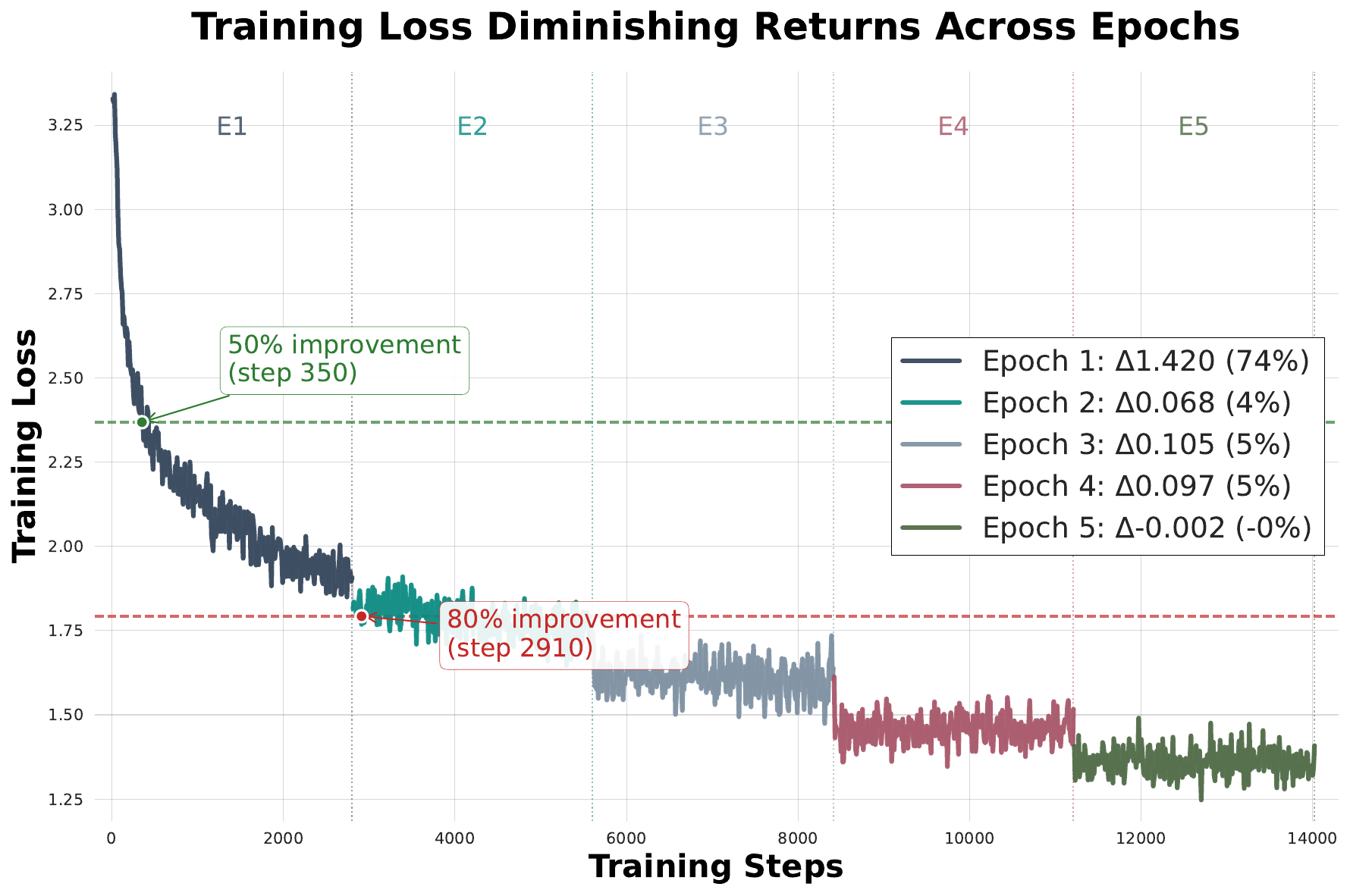}
        \caption{Empirical illustration of diminishing returns during LLM fine-tuning.}
    \label{fig:training_dynamics_demo}
    \vspace{-10pt}
\end{figure}

To validate this dynamic empirically, we fine-tuned an LLM \emph{LLama3.2-3B-instruct} on a medical QA dataset \emph{HealthCareMagic} using LoRA for five epochs. Figure \ref{fig:training_dynamics_demo} illustrates that initial epochs account for the majority of loss reductions (74\% in the first epoch). After
the first epoch, however, the learning process slowed down
significantly, with each subsequent epoch contributing less
than 5\% of the total reduction. By epoch five, the improvement becomes negligible.

\begin{figure}
    \centering
    \includegraphics[width=0.8\linewidth]{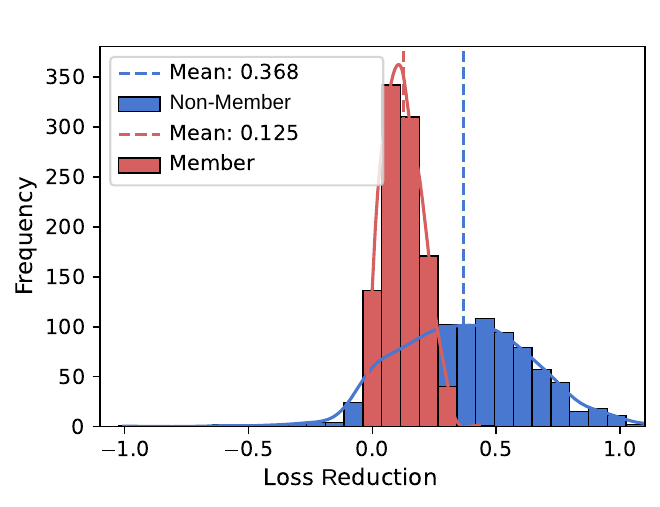}
    \caption{Fine-tuning with Members V.S. Non-Members}
    \label{fig:loss_difference}
    \vspace{-10pt}
\end{figure}

To further empirically validate how this diminishing-return behavior manifests as an optimization gap for membership inference, we conducted a controlled experiment comparing member and non-member samples. Specifically, we fine-tuned the base model \emph{LLama3.2-3B-instruct} for two epochs on the \emph{HealthCareMagic} dataset and then randomly selected 1,000 member samples and 1,000 non-member samples from the test dataset. We combined these samples, performed an additional short fine-tuning phase under identical training settings, and measured the per-sample loss reduction. As shown in Figure~\ref{fig:loss_difference}, non-members experience substantially larger loss drops (mean 0.368) than members (mean 0.125), with wider distributional spread, indicating much greater remaining optimization potential. This provides direct empirical support for the optimization-gap hypothesis.

While the optimization gap is a powerful membership signal, a practical challenge remains: \emph{How can an adversary measure this quantity in a black-box setting, where parameter updates are not permitted?} Real-world attackers can only observe model outputs—typically token-level log-likelihoods—through inference queries. They cannot perform additional training or compute gradients, rendering direct gap estimation infeasible. To address this, we propose a practical \emph{proxy method} for approximating the optimization gap in a black-box manner, which must satisfy the following criteria:
\begin{enumerate}[leftmargin=*]
\item \textbf{Training-Free}: The proxy must use inference queries only, without modifying model parameters.
\item \textbf{Efficient}: It should require only a small number of queries per sample.
\item \textbf{Sensitive}: It must reliably distinguish the optimization gaps of members vs. non-members.
\end{enumerate}

Our solution leverages the In-Context Learning (ICL) capabilities of LLMs: by strategically constructing probe contexts, we can simulate a fine-tuning-like update at inference time and observe the induced change in log-likelihood. This forms the basis of our proposed method, In-Context Probing (ICP), discussed in the next section.

\subsection{In-Context Probing as a Proxy for the Optimization Gap}
\label{sec:icp_proxy}
LLMs possess the remarkable capability known as In-Context Learning (ICL), enabling them to adapt and refine their outputs based solely on example context provided in the prompt, without modifying their parameters. Recent theoretical work interprets ICL as a form of implicit optimization, where the model internally simulates gradient-based adjustments in response to the provided context~\cite{dai2023WhycanGPT,akyureklearning}. Follow-up studies in data attribution support this interpretation, showing that carefully designed context perturbations can approximate gradient-based influence scores~\cite{jiao2025feasibilityincontextprobingdata}.

\paragraph{From True Optimization to In-Context Approximation}
Consider a true fine-tuning step in which the model $\mathcal{M}$ is trained on a sample $s=(x,y)$ to obtain $\mathcal{M}’$. The single-step optimization gain is captured by the log-likelihood (LL) improvement:
\begin{equation}
\Delta_{\text{LL}}(s) = LL(y \mid x; \mathcal{M}) - LL(y \mid x; \mathcal{M}')
\label{eq:opt_gap}
\end{equation}
where
\begin{equation}
LL(y \mid x; \mathcal{M}) = \sum_{t=1}^{L} \log p(y_t \mid x, y_{<t}; \mathcal{M})
\label{eq:baseline_ll}
\end{equation}

In ICL, we do not update model parameters. Instead, we prepend a probe context $C$ to the input $x$, creating a prompted input $C \oplus x$, and compute the conditioned LL:
\begin{equation}
LL(y \mid C \oplus x; \mathcal{M}) = \sum_{t=1}^{L} \log p(y_t \mid C \oplus x, y_{<t}; \mathcal{M})
\label{eq:probed_ll}
\end{equation}
If ICL indeed mimics gradient-based optimization, then $LL(y \mid C \oplus x; \mathcal{M})$ should approximate $LL(y \mid x; \mathcal{M}’)$ in~(\ref{eq:opt_gap}). In the next subsection, we empirically validate this approximation by measuring the correlation between ICL-induced loss changes and true gradient-based loss reductions.

\paragraph{In-Context Probing Score}
Motivated by this connection, we define the \textbf{In-Context Probing (ICP) score} for a probe $C$ as:
\begin{equation}
\text{ICP}_{\text{score}}(s, C) = LL(y \mid x; \mathcal{M}) - LL(y \mid C \oplus x; \mathcal{M})
\label{eq:icp_single_probe_score}
\end{equation}
which serves as a black-box approximation of the true optimization gain in~(\ref{eq:opt_gap}). The stronger the LL improvement induced by the probe, the larger the inferred optimization potential of sample $s$.

\subsection{Empirical Validation of ICP as an Optimization-Gap Proxy}
\label{sec:validation_proxy}

\begin{figure}
    \centering
    \includegraphics[width=0.85\linewidth]{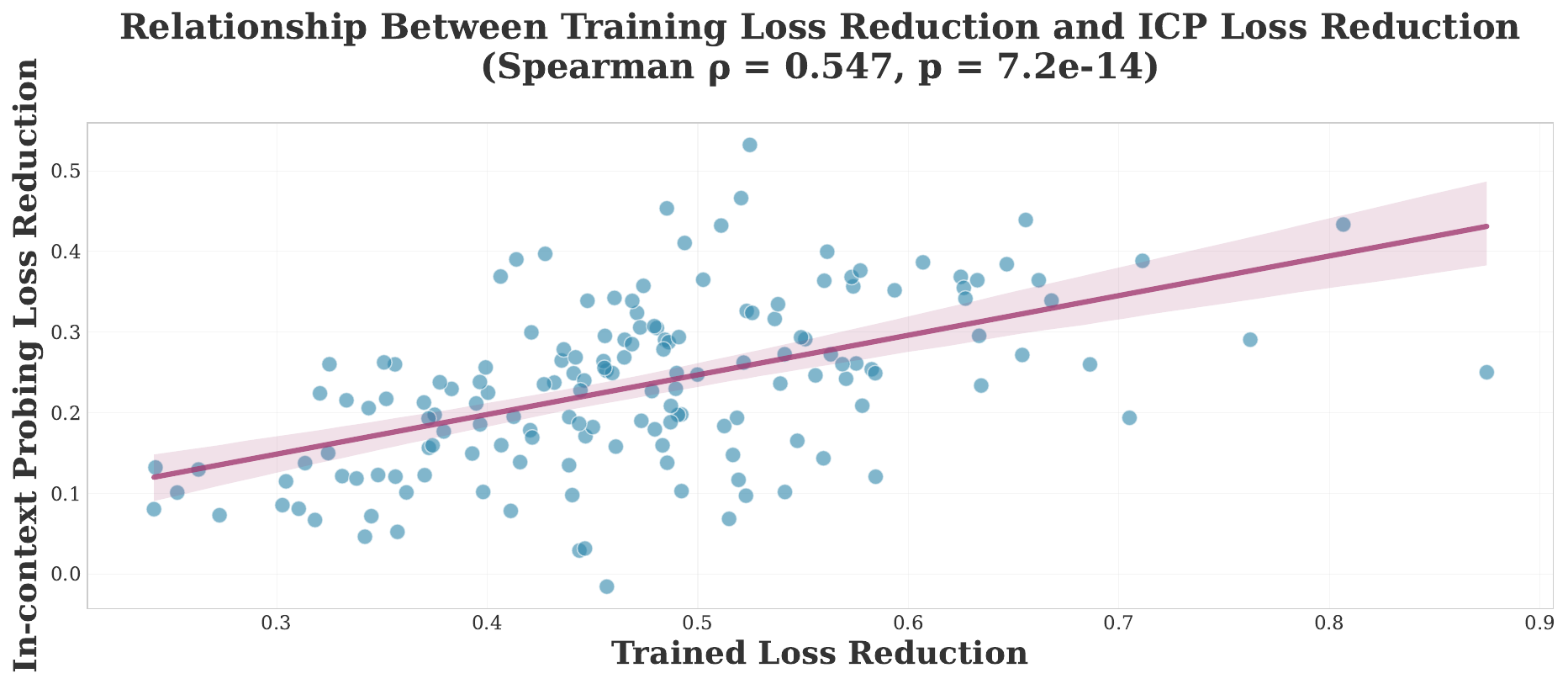}
    \caption{Correlation between actual single-step training loss reduction and the ICP-induced loss reduction.}
\label{fig:correlation_scatter}
\vspace{-10pt}
\end{figure}

\begin{figure*}
    \centering
    \includegraphics[width=0.85\linewidth]{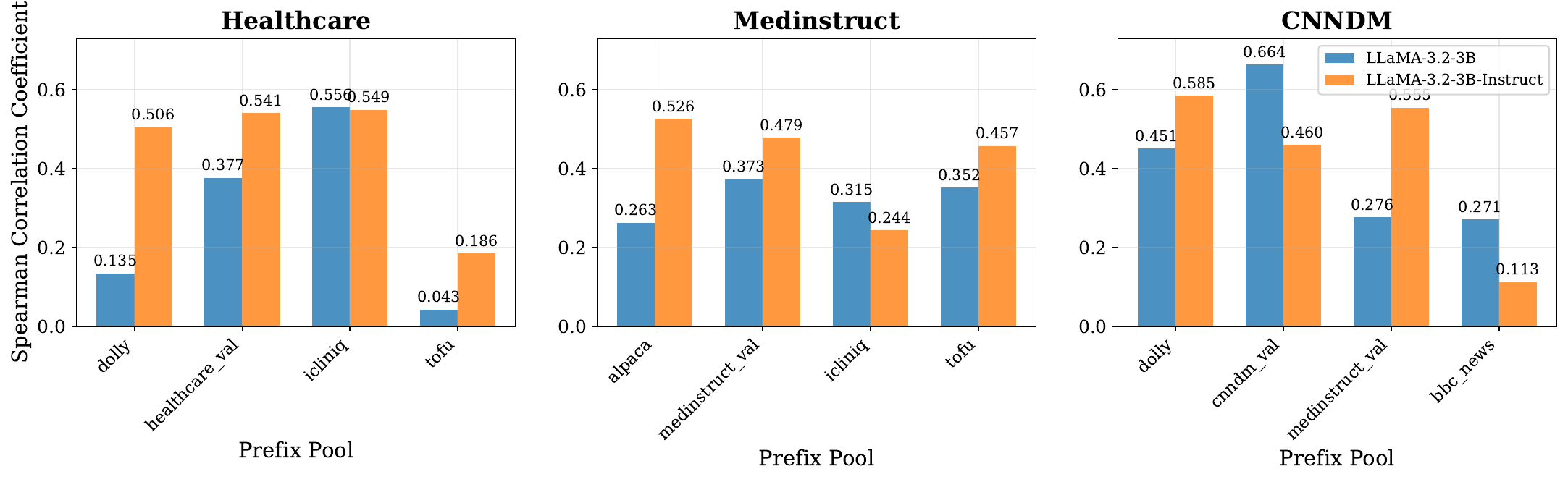}
    \caption{Impact of reference data (prefix pool) and model type on the fidelity of the ICP proxy, measured by Spearman correlation. The proxy demonstrates the highest effectiveness (strongest correlation) with instruction-tuned models and when the reference data closely aligns with the target task’s domain and semantics.}
    \label{fig:correlation_bar_charts}
    \vspace{-10pt}
\end{figure*}

To validate the effectiveness of ICP, we conduct controlled experiments and measure the correlation between the true loss reduction caused by an actual fine-tuning step and the loss reduction induced by our ICP approach. Specifically, we sample batches of 16 examples from the \emph{HealthcareMagic} dataset, perform an actual one-step gradient update to obtain the true loss reduction and compare this to the loss change induced by ICP probes. We generate ICP probes using the domain-aligned examples from the \emph{iCliniq} dataset. For each target sample, we select the top-20 most similar samples in embedding space (via cosine similarity), and choose the sample that maximizes the target sample's conditional likelihood as the prefixed probe context. Using these probe contexts, we compute the ICP-induced loss reduction.
Results from 10 independent runs (Figure~\ref{fig:correlation_scatter}) show that ICP achieves a statistically significant Spearman correlation of $\rho = 0.547$ (with $p$-value $7.2 \times 10^{-14}$), demonstrating that ICP provides a reliable approximation of the true Optimization Gap.

We further analyze how the choice of reference datasets used as prefix context pool and model architecture affects the fidelity of ICP. Specifically, we evaluate three datasets—Healthcare, MedInstruct, and CNN-DM—and two model variants, LLaMA-3.2-3B (base) and LLaMA-3.2-3B-Instruct (instruction-tuned), of which details can be found in Section \ref{ssec:expsetup}. For each dataset-model combination, we measure the Spearman correlation between the ICP-induced loss reduction and the actual single-step gradient-based loss reduction using a variety of prefix pools, ranging from domain-aligned datasets (e.g., \emph{iClinq} for Healthcare, \emph{medinstruct\_val} for MedInstruct, and \emph{cnndm\_val} for CNN-DM) to general-purpose datasets (e.g., \emph{dolly}, \emph{alpaca}) and unrelated datasets (e.g., \emph{tofu}, \emph{bbc\_news}).

As shown in Figure~\ref{fig:correlation_bar_charts}, instruction-tuned models (e.g., LLama-3.3-3B-instruct) overall exhibit higher correlations than their base models (e.g., LLama-3.3-3B), especially when paired with closely aligned reference datasets (e.g., iCliniq for HealthcareMagic). We attribute this to instruction tuning, strengthening a model’s inherent ICL mechanisms, enabling it to more accurately simulate a true optimization step in response to probing. Consequently, this improved simulation fidelity inadvertently amplifies the model’s susceptibility to our ICP-based attack.

The results also demonstrate that the quality of the proxy is dependent on the alignment between the reference dataset (prefix pool) and the target sample. Across all experiments, the highest correlation is achieved when using reference data from a dataset with high task and domain similarity (e.g., iCliniq or the in-distribution validation set for HealthcareMagic). General-purpose instruction datasets like Dolly and Alpaca yield a viable but weaker signal, while semantically and functionally unrelated datasets (e.g., using the TOFU dataset for the HealthcareMagic) cause the correlation to collapse, often becoming statistically insignificant. This confirms that semantically aligned probes are essential for eliciting meaningful optimization-like behavior.

Together, these experiments strongly validate our central hypothesis: ICP reliably approximates the Optimization Gap in a black-box setting. Moreover, they highlight that the effectiveness of ICP predictably depends on both the characteristics of the target model and the alignment of the probe context. These findings establish a principled and practical foundation for our proposed ICP-MIA framework.

\section{ICP-MIA: MIA via In-Context Probing}
\label{sec:ICP-MIA}
Building on the empirical validation of \textbf{In-Context Probing (ICP)} as an effective proxy for the Optimization Gap (Section~\ref{sec:validation_proxy}), we now describe \textbf{ICP-MIA}, which uses this proxy to perform membership inference in a black-box setting.

\subsection{Attack Formulation}
ICP-MIA estimates the optimization potential using the \emph{in-context probing score} defined earlier in Section~\ref{sec:icp_proxy}.
Given a sample $s=(x,y)$, ICP-MIA constructs probe contexts $C$ and evaluates the model on the original input $x$ and the probed input $C \oplus x$ using the ICP score, $\mathrm{ICP}_{\text{score}}(s,C)$, defined in Equation (\ref{eq:icp_single_probe_score}).
A strongly negative ICP score indicates a large LL improvement under the probe—suggesting the model could still benefit from additional training on the sample and thus indicating non-membership. Conversely, a small change in LL implies the sample has already been learned.

To maximize signal strength, ICP-MIA generates a set of $K$ candidate probe contexts, $\mathcal{C} = \{C_1, \ldots, C_K\}$. The most effective probe for a non-member is the one that elicits the largest LL improvement, which corresponds to the smallest $\mathrm{ICP}_{\text{score}}$. We therefore define the final membership score as:
\begin{equation}
\operatorname{Score}(s, \mathcal{C}) = \min_{C_j \in \mathcal{C}} \text{ICP}_{\text{score}}(s, C_j)
\label{eq:final_score}
\end{equation}
This score corresponds to the probe causing the largest negative change in log-likelihood. Our central hypothesis is that this score will distinguish members from non-members. Formally, we expect:
\begin{equation}
\mathbb{E}_{s \sim D_{\text{train}}}[\operatorname{Score}(s, \mathcal{C})] > \mathbb{E}_{s \sim D_{\text{test}}}[\operatorname{Score}(s, \mathcal{C})]
\label{eq:hypothesis}
\end{equation}

In other words, training samples yield higher scores (closer to zero) than unseen test samples on average. The rationale is that a member sample, having already been learned by the model, will not benefit much from any probe – its log-likelihood is already near optimal, so even the best probe only marginally increases confidence (resulting in a score near 0). In contrast, a non-member sample has significant unused optimization potential; a well-chosen probe can substantially increase the model’s confidence on that sample, leading to a strongly negative score. Following the common MIA framework in Section~\ref{ssec:problem}, a sample is predicted as a member if its score exceeds a threshold $\tau$, and as a non-member otherwise.

\subsection{Probe Context Construction}
\label{ssec:probeconstruct}
The effectiveness of ICP-MIA hinges on how we choose or generate the probe contexts $\mathcal{C}$. We introduce two complementary strategies: reference-data-based probing and self-perturbation probing, which offer different ways to elicit the hidden optimization gap for a sample.

\subsubsection{Reference-Data-Based Probing (ICP-MIA-Ref)}
In reference-data-based probing, we simulate a fine-tuning step using an auxiliary reference dataset $D_{\text{aux}}$ as prefix pool. For a given target sample $(x, y)$, we use embedding-based semantic retrieval to select $K$ input–output pairs from $D_{\text{aux}}$ that are semantically similar. 
Each retrieved pair provides a probe context $C_j$ that approximates how the model would behave if exposed to data resembling the target. See Appendix \ref{app:probe-ref-example} for an example.

\subsubsection{Self-Perturbation Probing (ICP-MIA-SP)}
The second strategy, self-perturbation probing, does not rely on any external data. Instead, in ICP-MIA-SP,  probes are directly generated from the target sample $s=(x,y)$, via two distinct approaches: generation-based perturbation and masking-based perturbation.

\paragraph{Generation-based Perturbation}
In this approach, we use an auxiliary language model as a generator $G$ to generate a set of $K$ variant responses, $\{y'_1, \dots, y'_K\}$, for the same input $x$. Each generated pair $(x, y'_j)$ then serves as a candidate context $C_j$ for target sample $(x,y)$. This method creates semantically relevant probes by generating plausible alternatives to the original response. The resulting candidate contexts is thus $\mathcal{C} = \{ (x, y'_j) \}_{j=1}^K$. By using such $(x, y'_j)$ as a prefix to the model, we probe how the target model’s confidence in the true answer $y$ changes when ``primed'' with a plausible alternative.

\paragraph{Masking-based Perturbation}
This approach creates probes by applying a binary mask $m \in \{0, 1\}^L$ to the original response $y$, where $L$ is the length of the output sequence. We replace token $y_t$ with a special \texttt{[MASK]} token if $m_t=1$, producing a partially masked sequence $y^{(m)}$. The resulting probe is $C_m = (x, y^{(m)})$. An example of a masking-based context probe is provided in Appendix \ref{app:example}. We generate a set of such masks using different strategies (described below), which form our candidate probe set $\mathcal{C}$ used in Equation (\ref{eq:final_score}).

\noindent\textbf{Random Masking.} This method generates masks by randomly selecting $\lfloor pL \rfloor$ token positions to mask, where $p$ is the masking ratio. Random masking tests the model’s robustness to missing information: if the model’s log-likelihood on $y$ is largely unaffected by removing arbitrary tokens, it suggests that the model already has a strong internal representation of the sample (as expected for members).
    
\noindent\textbf{LL-based Masking.} This method uses the model's own token-level LL (i.e., Log-Likelihood), $\ell_t = \log p(y_t \mid x, y_{<t}; \mathcal{M})$, to select which tokens to mask. We mask either the $\lfloor pL \rfloor$ tokens with the lowest $\ell_t$ (high-information, ``surprising'' tokens), referred to as Min-K\% Masking, or the highest $\ell_t$ (predictable, low-information tokens), referred to as Max-K\% Masking.
Masking these two types of positions allows us to probe the model’s sensitivity to removing either memorized content or routine patterns.

\section{Evaluation}
We conduct extensive experiments to evaluate our proposed ICP-MIA framework, considering both the reference-based (Ref) and self-perturbation (SP) variants. We first present our experimental setup, including models, datasets, and metrics, followed by comprehensive results across multiple settings.

\subsection{Experimental Setup}
\label{ssec:expsetup}

\begin{table*}[ht]
\centering
\caption{Comparison of MIA Methods across Different Models and Datasets}
\label{tab:main_results}
\setlength{\tabcolsep}{4pt} 
\begingroup
\renewcommand{\arraystretch}{1.2}
\definecolor{lightgray}{gray}{0.9}
\definecolor{lightblue}{RGB}{173,216,230}

\resizebox{\textwidth}{!}{%
\begin{tabular}{l|ccc|ccc|ccc}
\toprule
\multirow{3}{*}{\textbf{MIA Method}} & \multicolumn{9}{c}{\textbf{AUC}} \\
\cmidrule{2-10}
 & \multicolumn{3}{c|}{\textbf{LLaMA-3.2-3B-Instruct}} & \multicolumn{3}{c|}{\textbf{LLaMA-3.2-3B}} & \multicolumn{3}{c}{\textbf{Pythia-2.8B-deduped}} \\
\cmidrule(lr){2-4} \cmidrule(lr){5-7} \cmidrule(lr){8-10}
 & \textbf{Healthcare} & \textbf{MedInstruct} & \textbf{CNN-DM} & \textbf{Healthcare} & \textbf{MedInstruct} & \textbf{CNN-DM} & \textbf{Healthcare} & \textbf{MedInstruct} & \textbf{CNN-DM} \\
\midrule

\texttt{Bag of Words} & 0.485 & 0.512 & 0.502 & 0.491 & 0.493 & 0.536 & 0.501 & 0.497 & 0.516 \\
\texttt{Loss Attack} & 0.770 & 0.907 & 0.929 & 0.708 & 0.904 & 0.885 & 0.701 & 0.849 & 0.851 \\
\texttt{Zlib} & 0.765 & \underline{0.921} & \underline{0.932} & 0.703 & \underline{0.917} & 0.888 & 0.694 & \underline{0.866} & \underline{0.856} \\
\texttt{Min-K\%} & 0.837 & 0.907 & 0.930 & 0.763 & 0.908 & \underline{0.890} & 0.777 & 0.865 & \textbf{0.859} \\
\texttt{Min-K\%++} & 0.798 & 0.810 & 0.861 & 0.710 & 0.787 & 0.794 & 0.727 & 0.758 & 0.760 \\
\texttt{Neighborhood} & 0.669 & 0.556 & 0.661 & 0.614 & 0.535 & 0.621 & 0.635 & 0.527 & 0.627 \\
\texttt{Recall} & \underline{0.847} & 0.899 & 0.930 & \underline{0.780} & 0.908 & 0.884 & 0.768 & 0.854 & 0.820 \\
\midrule
\cellcolor{lightgray}\texttt{ICP-MIA-Ref} & \cellcolor{lightgray}0.827 & \cellcolor{lightgray}0.838 & \cellcolor{lightgray}0.890 & \cellcolor{lightgray}\textbf{0.842} & \cellcolor{lightgray}0.775 & \cellcolor{lightgray}0.837 & \cellcolor{lightgray}\underline{0.850} & \cellcolor{lightgray}0.746 & \cellcolor{lightgray}0.706 \\
\cellcolor{lightgray}\texttt{ICP-MIA-SP} & \cellcolor{lightgray}\textbf{0.942} & \cellcolor{lightgray}\textbf{0.959} & \cellcolor{lightgray}\textbf{0.965} & \cellcolor{lightgray}0.763 & \cellcolor{lightgray}\textbf{0.977} & \cellcolor{lightgray}\textbf{0.927} & \cellcolor{lightgray}\textbf{0.853} & \cellcolor{lightgray}\textbf{0.882} & \cellcolor{lightgray}0.845 \\
\midrule
\texttt{Reference Attack (Base)$^*$} & 0.796 & 0.885 & 0.925 & 0.736 & 0.878 & 0.871 & 0.717 & 0.871 & 0.856 \\
\texttt{Reference Attack (Ref)$^*$} & 0.870 & 0.902 & 0.971 & 0.817 & 0.898 & 0.937 & 0.799 & 0.891 & 0.919 \\
\texttt{SPV-MIA}$^*$ & 0.781 & 0.946 & 0.974 & 0.725 & 0.932 & 0.959 & 0.713 & 0.869 & 0.938 \\

\bottomrule
\end{tabular}%
}
\resizebox{\textwidth}{!}{%
\begin{tabular}{l|ccc|ccc|ccc}
\toprule
\multirow{3}{*}{\textbf{MIA Method}} & \multicolumn{9}{c}{\textbf{TPR@1\%FPR}} \\
\cmidrule{2-10}
 & \multicolumn{3}{c|}{\textbf{LLaMA-3.2-3B-Instruct}} & \multicolumn{3}{c|}{\textbf{LLaMA-3.2-3B}} & \multicolumn{3}{c}{\textbf{Pythia-2.8B-deduped}} \\
\cmidrule(lr){2-4} \cmidrule(lr){5-7} \cmidrule(lr){8-10}
 & \textbf{Healthcare} & \textbf{MedInstruct} & \textbf{CNN-DM} & \textbf{Healthcare} & \textbf{MedInstruct} & \textbf{CNN-DM} & \textbf{Healthcare} & \textbf{MedInstruct} & \textbf{CNN-DM} \\
\midrule

\texttt{Bag of Words} & 0.008 & 0.014 & 0.004 & 0.014 & 0.010 & 0.002 & 0.003 & 0.002 & 0.008 \\
\texttt{Loss Attack} & 0.042 & 0.266 & 0.088 & 0.028 & 0.256 & 0.034 & 0.020 & 0.168 & 0.020 \\
\texttt{Zlib} & 0.036 & 0.096 & 0.058 & 0.034 & 0.076 & 0.042 & 0.006 & 0.028 & 0.034 \\
\texttt{Min-K\%} & 0.046 & \underline{0.288} & 0.104 & 0.024 & \underline{0.412}& 0.090 & \underline{0.022} & \underline{0.176} & 0.046 \\
\texttt{Min-K\%++} & 0.034 & 0.090 & 0.116 & 0.004 & 0.060 & 0.028 & 0.016 & 0.036 & 0.010 \\
\texttt{Neighborhood} & 0.032 & 0.008 & 0.012 & 0.014 & 0.010 & 0.008 & 0.018 & 0.010 & 0.006 \\
\texttt{Recall} & 0.024 & 0.133 & \underline{0.195} & 0.014 & 0.044 & \underline{0.096} & 0.020 & 0.162 & \underline{0.108} \\
\midrule
\cellcolor{lightgray}\texttt{ICP-MIA-Ref} & \cellcolor{lightgray}\underline{0.084}& \cellcolor{lightgray}0.044 & \cellcolor{lightgray}0.020 & \cellcolor{lightgray}\textbf{0.140} & \cellcolor{lightgray}0.018 & \cellcolor{lightgray}0.062 & \cellcolor{lightgray}\underline{0.110} & \cellcolor{lightgray}0.074 & \cellcolor{lightgray}0.022 \\
\cellcolor{lightgray}\texttt{ICP-MIA-SP} & \cellcolor{lightgray}\textbf{0.172} & \cellcolor{lightgray}\textbf{0.326} & \cellcolor{lightgray}\textbf{0.518} & \cellcolor{lightgray}\underline{0.070} & \cellcolor{lightgray}\textbf{0.538} & \cellcolor{lightgray}\textbf{0.418} & \cellcolor{lightgray}\textbf{0.122} & \cellcolor{lightgray}\textbf{0.270} & \cellcolor{lightgray}\textbf{0.144} \\
\midrule
\texttt{Reference Attack (Base)$^*$} & 0.018 & 0.078 & 0.354 & 0.026 & 0.082 & 0.270 & 0.016 & 0.244 & 0.191 \\
\texttt{Reference Attack (Ref)$^*$} & 0.012 & 0.166 & 0.388 & 0.010 & 0.142 & 0.412 & 0.010 & 0.414 & 0.390 \\
\texttt{SPV-MIA}$^*$ & 0.034 & 0.486 & 0.602 & 0.036 & 0.608 & 0.440 & 0.020 & 0.374 & 0.531 \\

\bottomrule
\end{tabular}%
}

\caption*{\small Note: \textbf{Bold} indicates the best overall performance, and \underline{underline} indicates the second best performance among reference-free methods. For reference-based methods, Reference Attack (Base) uses the pretrained model itself as the reference, while Reference Attack (Ref) and SPV-MIA fine-tune a reference model on a held-out in-distribution split from the same dataset used to fine-tune the target model, giving them their strongest possible setting.}
\endgroup
\vspace{-10pt}
\end{table*}

\begin{figure*}[htpb]
    \centering
    
    \begin{subfigure}{0.9\textwidth}
        \centering
        \includegraphics[width=\textwidth]{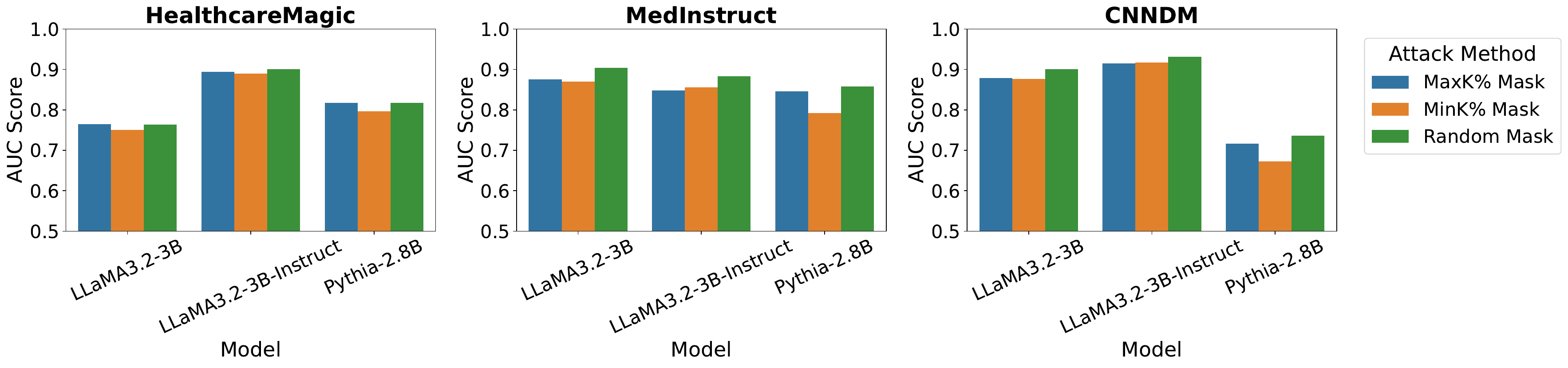}
        \label{fig:mask_auc}
    \end{subfigure}
        \vspace{-10pt}

    \begin{subfigure}{0.9\textwidth}
        \centering
        \includegraphics[width=\textwidth]{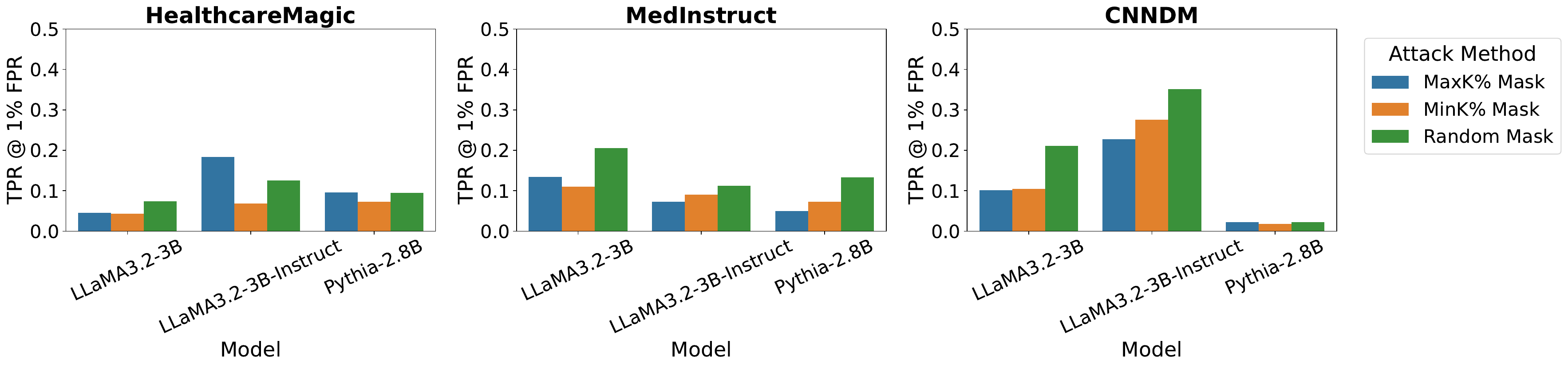}
        \label{fig:mask_tpr}
    \end{subfigure}
        \vspace{-10pt}
    \caption{Comparison of Different Masking-based Probing Strategies}
    \label{fig: maskstrategies}
\end{figure*}

\begin{figure*}[htbp]
    \centering
    \includegraphics[width=0.9\linewidth]{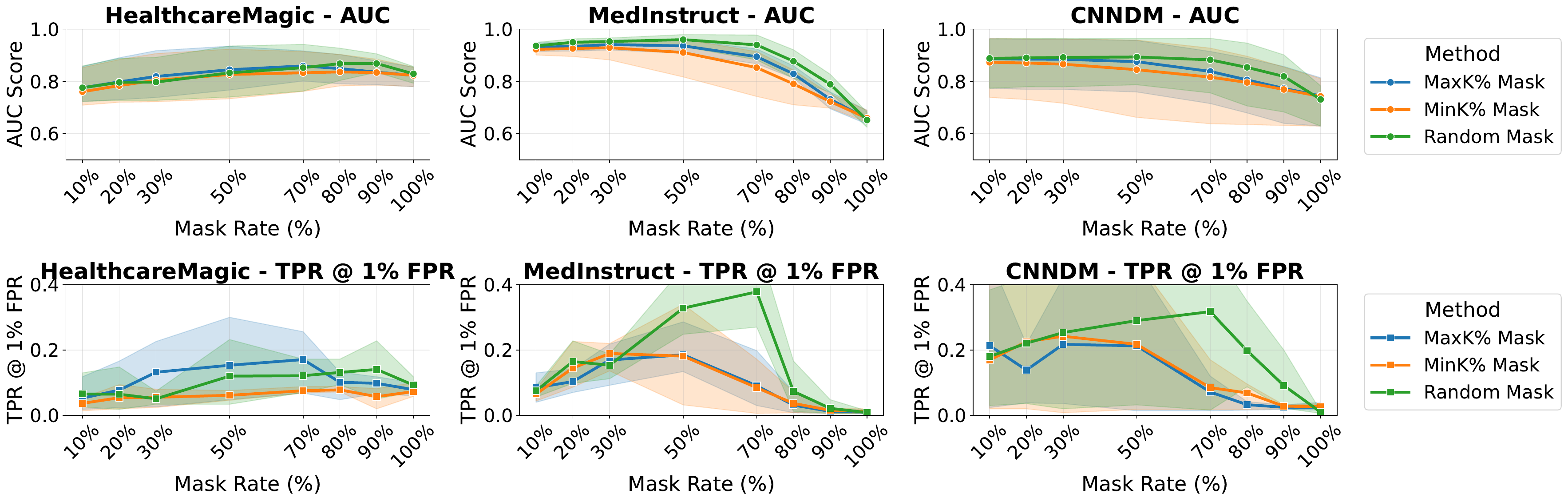}
    \caption{Mask-rate ablation for the masking-based ICP-MIA-SP variant. }
    \label{fig:mask_rate_ablation}
    \vspace{-10pt}
\end{figure*}

\noindent\textbf{Target Models.} We selected a suite of publicly available LLMs to ensure broad applicability and observe performance across different architectures and scales. Specifically, our experiments utilized Pythia-2.8B-deduped~\cite{biderman2023pythiasuiteanalyzinglarge}, Llama-3.2-3B, and Llama-3.2-3B-instruct~\cite{grattafiori2024llama}. These models allow us to examine attack robustness across a mix of non-instruction-tuned and instruction-tuned architectures.

\noindent\textbf{Datasets.} We primarily focus on three datasets: HealthcareMagic~\cite{li2023chatdoctor}, CNN-DM~\cite{hermann2015teaching}, and AlpacaCare-MedInstruct-52k~\cite{zhang2023alpacare}. For each dataset, 80\% of the data was used for fine-tuning the target LLMs (member set), and the remaining 20\% is split evenly into validation and test sets. Non-member samples were drawn exclusively from the test set to ensure that both members and non-members originate from the same data distribution. 
This setup avoids membership leakage due to distributional shift artifacts~\cite {duan2024membership} and ensures a fair MIA evaluation that reflects only memorization from the SFT process. We also evaluate ICP-MIA on two supplementary datasets—XSum~\cite{xsum} and AG News~\cite{zhang2015characterlevel}—used in prior MIA work. The corresponding results, which exhibit similar trends, are included in Appendix~\ref{app:extradataset} due to space constraints.

\noindent\textbf{Baseline.} We compare with seven state-of-the-art MIAs. To control for potential distribution shift, we include \texttt{Bag of Words}~\cite{meeus2024sokmembershipinferenceattacks} as the blind baseline, which uses a random forest classifier on bag of words features to determine membership. Its performance should be close to random guessing. \texttt{Loss Attack}~\cite{yeom2018privacy} uses loss of input as membership score. \texttt{Zlib}~\cite{carlini2021extracting} normalizes the loss using Zlib entropy and uses it as membership score. \texttt{Min-K\%}~\cite{shidetecting} averages the log-likelihood over the top K\% lowest-probability tokens of input as its membership score. \texttt{Min-K\%++}~\cite{zhang2024min} method further refines this signal by identifying local maxima in the model's conditional distributions. 
\texttt{Recall}\cite{xie-etal-2024-recall} computes the ratio between a sample’s log-likelihood under a non-member prefix and its original log-likelihood. \texttt{Neighborhood}\cite{mattern2023neighborhood} scores a sample by comparing its log-likelihood to that of its perturbed neighbors. Finally, \texttt{SPV-MIA}\cite{fu2024practicalmembershipinferenceattacks} measures the probability gap between a sample and its symmetric semantic neighbors, calibrated using a self-prompted reference model.

\noindent\textbf{Metrics.}
We use \emph{AUC} and \emph{TPR@Low FPR} as our evaluation metrics. Following standard practice~\cite{carlini2022membership}, we specifically report TPR@1\%FPR to reflect real-world attack scenarios where false positives are costly.

\noindent\textbf{Experimental Details.}
Unless otherwise noted, all models are fully fine-tuned for two epochs. This setup ensures that the models are sufficiently adapted to the downstream tasks, creating a realistic scenario for evaluating membership inference vulnerability. By default, we use $K=5$ candidate probes for ICP-MIA-SP and $K=10$ for ICP-MIA-Ref. Detailed hyperparameter configurations for the fine-tuning process are provided in Appendix~\ref{app:Experiment_Details}.

For \texttt{ICP-MIA-Ref}, we use Dolly-15k as the reference dataset across all experiments—a general-purpose instruction-following dataset that demonstrates our method's effectiveness without requiring domain-specific or distribution-matched data. We retrieve semantically similar probes using embeddings from sentence-transformers/all-MiniLM-L6-v2. 

For \texttt{ICP-MIA-SP}, we generate diverse response variants using four state-of-the-art LLMs: Llama-3.3-70B-Instruct~\cite{grattafiori2024llama}, Qwen2-72B-Instruct~\cite{qwen2025qwen25technicalreport}, Mixtral-8x22B-Instruct~\cite{jiang2024mixtralexperts}, and GPT-4.1-mini~\cite{openai_gpt4_1mini_2025}. Generation prompts are provided in Appendix~\ref{app:prompt_perturbation_generation}. 

\subsection{Main Results}
Our main results are summarized in Table~\ref{tab:main_results}. We highlight several key observations:

\noindent\textbf{1. ICP-MIA consistently outperforms
existing methods.}
Across most experimental configurations, our ICP-MIA framework outperforms all reference-free methods. In particular, \texttt{ICP-MIA-SP} consistently ranks among the top-performing attacks: on LLaMA-3.2-3B-Instruct, it achieves AUC scores of 0.942, 0.959, and 0.965 on Healthcare, MedInstruct, and CNN-DM, respectively—substantially surpassing all reference-free baselines. 

Moreover, under LLaMA models, ICP-MIA often matches or even exceeds the performance of reference-based methods despite requiring no reference-model training. These results highlight the strength of the Optimization Gap as a membership signal. A notable exception occurs under Pythia-2.8B-deduped, especially with CNN-DM, where \texttt{ICP-MIA-SP} achieves a TPR@1\%FPR of 0.144, substantially underperforming \texttt{SPV-MIA}. We attribute this gap to differences in models’ ability to exploit contextual information. LLaMA-3.2 models are optimized for instruction-following and in-context reasoning, whereas Pythia is not. Because ICP-MIA relies on strong ICL capabilities to simulate fine-tuning behavior, weaker ICL leads to less accurate Optimization Gap estimates and reduced attack performance.

\noindent\textbf{2. The advantage of ICP-MIA is most evident in challenging, realistic scenarios.} The results reveal that HealthcareMagic is the most difficult dataset for MIA, with most baseline methods yielding low AUC and TPR. Yet, in this challenging scenario, our method's advantage is most stark. For example, on the Pythia model with HealthcareMagic, \texttt{ICP-MIA-SP} achieves an AUC of 0.853 and a TPR of 0.122, while the next-best baseline (\texttt{Reference Attack(Ref)}) only reaches 0.799 AUC and a far lower 0.010 TPR. This demonstrates that ICP-MIA captures a more reliable signal, which is less susceptible to the data characteristics that confound traditional loss-based attacks.

\noindent\textbf{3. Instruction-tuning tends to increase membership vulnerability.} By comparing the results for LLaMA-3.2-3B-Instruct against its base model, LLaMA-3.2-3B, we observe that instruction-tuning increases the effectiveness of our ICP-based attack. On CNN-DM, the AUC of \texttt{ICP-MIA-SP} jumps from 0.927 to 0.965, while on HealthcareMagic it rises from 0.763 to 0.942. This suggests that instruction-tuning enhances the model's ability to leverage contextual information in prompts, making the Optimization Gap between members and non-members more pronounced. This conclusion is further supported by Figure~\ref{fig:correlation_bar_charts}, which shows higher correlation between true loss reduction and our ICP-based proxy score for instruction-tuned models.

\subsection{Evaluating Masking-based Probing for ICP-MIA-SP}
\label{sec:eval_masking_strategies}

We conduct a comparative analysis of our masking-based probing strategies to understand which type of perturbation most effectively exposes a membership signal. Specifically, we compare \textit{Random Masking} against two LL-based methods: \textit{Min-K\% Masking} (targeting surprising tokens) and \textit{Max-K\% Masking} (targeting predictable tokens). As shown in Figure~\ref{fig: maskstrategies}, \textit{Random Masking} consistently outperforms the two LL-based methods across nearly all experimental settings.

We argue that the success of \textit{Random Masking} stems from the difficulty of selecting an effective probe. Finding a mask configuration that reliably exposes the optimization gap is non-trivial—simple heuristics like targeting high-LL or low-LL tokens may not consistently identify the most revealing perturbations. LL-based methods produce only a \textit{single deterministic probe} per sample based on such heuristics, limiting their ability to discover the optimal configuration. In contrast, \textit{Random Masking} generates \textit{multiple independent probes} and selects the one producing the strongest signal. By sampling from a broader space of mask configurations, this approach has a higher probability of finding an effective probe for each sample. These results show that random multi-probing outperforms LL-based heuristics, highlighting that diversity from multiple trials is more effective than deterministic selection.

To determine the optimal perturbation magnitude for our masking-based probes, we performed an ablation study on the mask rate, $p$, which represents the percentage of tokens in the response that are masked. We evaluate $p$ from 10\% to 100\%. Figure~\ref{fig:mask_rate_ablation} shows a consistent non-monotonic trend: attack performance generally increases with the mask rate, reaches an optimal point, and then degrades.

\begin{figure}[htpb]
    \centering
    \includegraphics[width=0.95\linewidth]{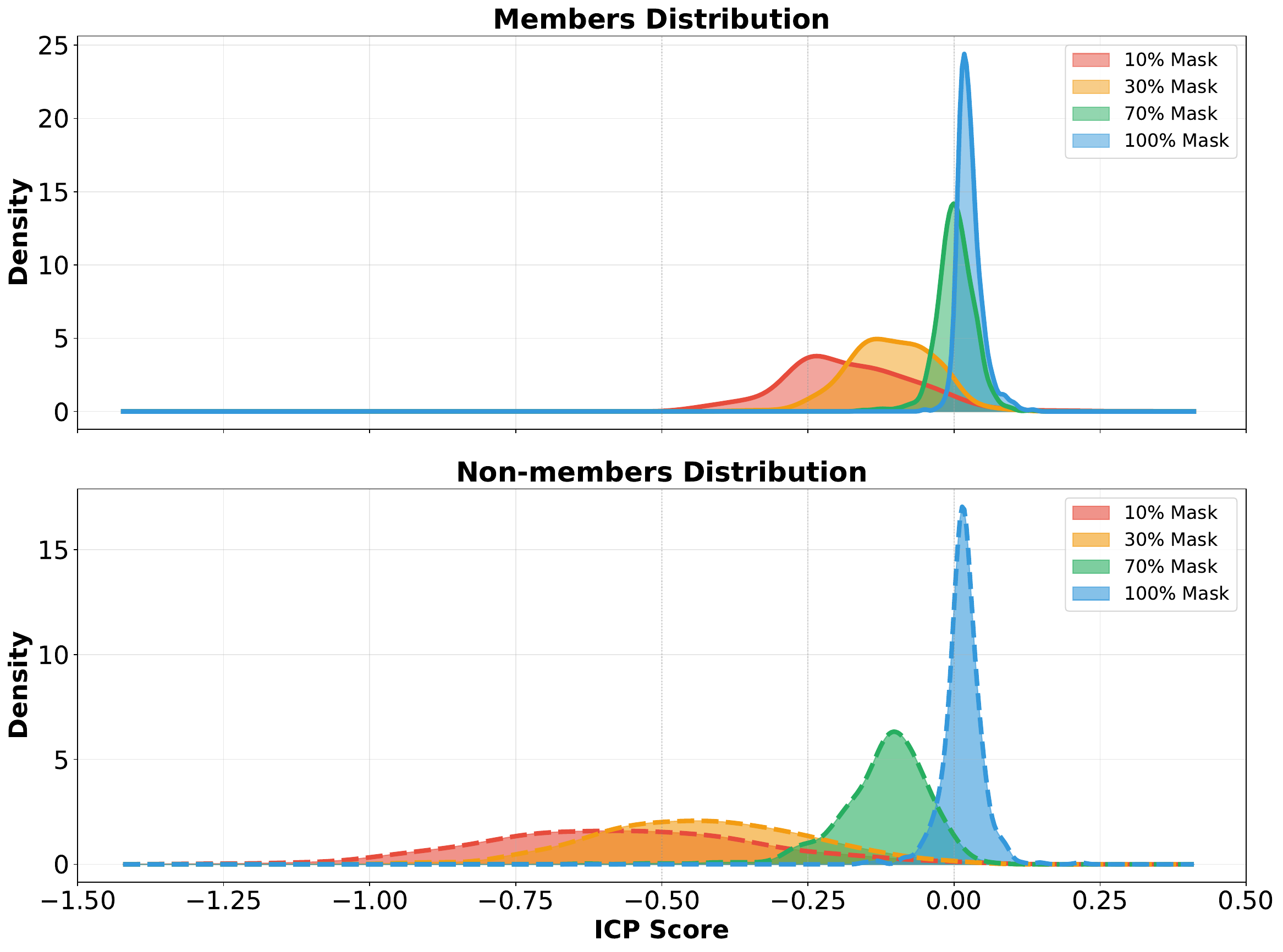}
    \caption{ICP scores distribution under different masking percentages.}
    \label{fig:icp_distribution}
    \vspace{-10pt}
\end{figure}

We analyze these single-peaked curves by examining four distinct ranges of the mask rate. As shown in Figure~\ref{fig:icp_distribution}, at very low mask rates ($p < 30\%$), the attack is ineffective because the probe retains too much information. The probe preserves much of the original answer's semantic content which boosts confidence for both member and non-member samples, resulting in poor class separation and weak attack performance. As the mask rate increases to a moderate range (approximately 30\%--70\%), attack performance peaks. In this range, enough information has been removed to challenge the model's understanding, forcing it to rely on internalized knowledge. For member samples, which have been memorized during fine-tuning, the partial context provides only marginal improvement. For non-members, however, the model gains substantial benefit from the contextual hints, as the information is novel. This asymmetry in log-likelihood improvement maximizes the membership signal. However, beyond this optimal point, performance degrades. At excessively high mask rates ($p > 80\%$), the probe becomes information-sparse, consisting almost entirely of \texttt{[MASK]} tokens. Such probes offer negligible guidance to the model for either class, causing the membership signal to vanish. This degradation is particularly problematic in high-precision scenarios, as evidenced by the sharp drop in TPR@1\%FPR at high mask rates.

\subsection{Evaluating Generation-based Probing for ICP-MIA-SP}
To evaluate the impact of the perturbation generator $G$ on ICP-MIA-SP, we compare three open-source models and GPT-4.1-mini in Table~\ref{tab:ablation_genmodel}. The table shows that Qwen-2.5-72B-Instruct delivers the best results, outperforming even the GPT-4.1-mini. This indicates that our method does not depend on expensive APIs, as high-quality open-source generators can provide comparable or even superior performance. In our ablation study, we also vary the sampling temperature for each open-source model from 0.2 to 1.0 and compute the average AUC and TPR@1\%FPR. The results show that temperature has little impact on AUC, but TPR@1\% FPR shows noticeably higher variance and is more sensitive to temperature changes. Nevertheless, Qwen-2.5-72B-Instruct consistently demonstrates strong overall performance across both metrics. Figures illustrating the effect of temperature are presented in Appendix Figure~\ref{fig:temperature_ablation}.

\begin{table*}[htbp]
\centering
\caption{ICP-MIA-SP Performance with Different Generator Models}
\label{tab:ablation_genmodel}
\begin{tabular}{l|l|cc|cc|cc}
\hline
\multirow{2}{*}{\textbf{Generator}} & \multirow{2}{*}{\textbf{Dataset}} & \multicolumn{2}{c|}{\textbf{Llama3.2-3B}} & \multicolumn{2}{c|}{\textbf{Llama3.2-3B-Instruct}} & \multicolumn{2}{c}{\textbf{Pythia-2.8B-Deduped}} \\
\cline{3-8}
 &  & \textbf{AUC} & \textbf{TPR@1\%FPR} & \textbf{AUC} & \textbf{TPR@1\%FPR} & \textbf{AUC} & \textbf{TPR@1\%FPR} \\
\hline
\hline
\multirow{3}{*}{Qwen2.5-72B-Instruct} & CNN-DM & \textbf{0.938} & \textbf{0.394} & 0.968& 0.533 & 0.750 & 0.020 \\
 & MedInstruct & \textbf{0.953}& 0.176 & \textbf{0.961} & \textbf{0.283} & \textbf{0.903} & 0.077 \\
 & HealthcareMagic & 0.792 & \textbf{0.065} & \textbf{0.905} & \textbf 0.098 & \textbf{0.797} & \textbf{0.123} \\
\hline
\multirow{3}{*}{Llama-3.3-70B-Instruct} & CNN-DM & 0.928 & 0.273 & 0.965 & 0.558 & 0.762 & \textbf{0.025} \\
 & MedInstruct & 0.898 & 0.122 & 0.911 & 0.170 & 0.857 & 0.087 \\
 & HealthcareMagic & 0.748 & 0.046 & 0.872 & 0.094 & 0.765 & 0.110 \\
\hline
\multirow{3}{*}{Mixtral-8x22B-Instruct} & CNN-DM & 0.933 & 0.382 & 0.966 & \textbf{0.640} & 0.748 & 0.018 \\
 & MedInstruct & 0.940 & 0.108 & 0.947 & 0.108 & 0.901 & \textbf{0.099} \\
 & HealthcareMagic & 0.758 & 0.060 & 0.898 & 0.057 & 0.787 & 0.080 \\
\hline
\multirow{3}{*}{GPT-4.1-mini} & CNN-DM & 0.920 & 0.124 & \textbf{0.969} & 0.340 & \textbf{0.856} & 0.010 \\
 & MedInstruct & 0.940 & \textbf{0.412} & 0.946 & 0.260 & 0.876 & 0.054 \\
 & HealthcareMagic & \textbf{0.864} & 0.042 & 0.850 & \textbf{0.144} & 0.735 & 0.016 \\
\hline
\end{tabular}
\end{table*}

\subsection{The impact of Public Dataset Selection For ICP-MIA-Ref} 
To investigate how the choice of public datasets affects the performance of \texttt{ICP-MIA-Ref}, we tested it on the Healthcaremagic dataset using three reference datasets with varying alignment. Alpaca \cite{taori2023stanford}, a general-purpose instruction-following dataset, provides broad and semantically diverse examples. iCliniq~\cite{li2023chatdoctor}, a medical question-and-answer dataset, closely matches Healthcaremagic in both task type and semantic content, making it highly relevant. TOFU~\cite{maini2024tofu}, a QA dataset generated by fictitious authors, shares the question-answer format but differs significantly in semantic content. This setup allows us to systematically examine how semantic similarity and task alignment between the public dataset and the fine-tuning data impact MIA effectiveness.

\begin{table}[!ht]
\centering
\caption{Public Dataset Impact on ICP-MIA-Ref}
\label{tab:icp_mia_ref_ablation}
\scriptsize %
\setlength{\tabcolsep}{3pt} %
\begin{tabular}{@{}l|cc|cc|cc|cc@{}}
\toprule
\multirow{2}{*}{\textbf{Model}} & \multicolumn{2}{c|}{\textbf{Alpaca}} & \multicolumn{2}{c|}{\textbf{iCliniq}} & \multicolumn{2}{c|}{\textbf{TOFU}} & \multicolumn{2}{c@{}}{\textbf{Validation}} \\
\cmidrule(lr){2-3} \cmidrule(lr){4-5} \cmidrule(lr){6-7} \cmidrule(lr){8-9}
 & \textbf{AUC} & \textbf{TPR} & \textbf{AUC} & \textbf{TPR} & \textbf{AUC} & \textbf{TPR} & \textbf{AUC} & \textbf{TPR} \\
\midrule
LLaMA-3.2-3B & 0.813 & 0.072 & \textbf{0.873} & \underline{0.188} & 0.743 & 0.020 & \underline{0.864} & \textbf{0.320} \\
LLaMA-3.2-3B-Instruct & 0.819 & \underline{0.168} & \underline{0.857} & 0.112 & 0.821 & 0.146 & \textbf{0.943} & \textbf{0.194} \\
Pythia-2.8B-Deduped & 0.817 & 0.042 & \underline{0.830} & \textbf{0.122} & 0.825 & \underline{0.100} & \textbf{0.875} & 0.074 \\
\bottomrule
\end{tabular}
\end{table}

Table \ref{tab:icp_mia_ref_ablation} demonstrates that both task structure and semantic alignment independently influence ICP-MIA performance. iCliniq, matching HealthcareMagic in task format (QA) and domain (medical), achieves strong results (AUC up to 0.873). TOFU occasionally outperforms Alpaca despite semantic mismatch, indicating that task structural consistency provides benefits beyond semantic similarity. The validation dataset from the same source (Healthcaremagic) yields optimal performance (e.g., AUC=0.943), confirming that complete distributional alignment maximizes attack effectiveness. These findings validate our hierarchical approach for reference selection: prioritize datasets aligned in both task and semantics, followed by those partially aligned alternatives.

\subsection{Impact of Parameter-Efficient Fine-tuning Methods}

To evaluate the robustness of our framework in practical scenarios, we investigate the impact of different Parameter-Efficient Fine-tuning methods on MIA. We fine-tuned a base model using Low-Rank Adaptation (LoRA) with varying capacities (rank $r=32$ and $r=64$) and its quantized variant, QLoRA ($r=64$, 4-bit and 8-bit ). As demonstrated in Table~\ref{tab:peft_results}, our ICP-MIA methods, particularly \texttt{ICP-MIA-Ref}, consistently outperform baseline attacks across all PEFT configurations. This result validates that the underlying Optimization Gap signal, which our method is designed to measure, persists even when parameter updates are constrained.

\begin{table}[ht]
\centering
\caption{MIA Across Different PEFT Methods}
\label{tab:peft_results}
\scriptsize
\setlength{\tabcolsep}{3pt}
\begin{tabular}{@{}l|cc|cc|cc|cc@{}}
\toprule
\multirow{2}{*}{\textbf{MIA Method}} & \multicolumn{2}{c|}{\textbf{LoRA (r=32)}} & \multicolumn{2}{c|}{\textbf{LoRA (r=64)}} & \multicolumn{2}{c|}{\textbf{QLoRA (4bit)}} & \multicolumn{2}{c@{}}{\textbf{QLoRA (8bit)}} \\
\cmidrule(lr){2-3} \cmidrule(lr){4-5} \cmidrule(lr){6-7} \cmidrule(lr){8-9}
& AUC & TPR & AUC & TPR & AUC & TPR & AUC & TPR \\
\midrule
\texttt{Loss Attack} & 0.623 & 0.030 & 0.661 & 0.022 & 0.646 & 0.024 & 0.653 & 0.028 \\
\texttt{Zlib} & 0.642 & 0.025 & 0.656 & 0.022 & 0.635 & 0.023 & 0.647 & 0.026 \\
\texttt{Min-K\%} & 0.645 & 0.026 & 0.693 & 0.028 & 0.678 & 0.032 & 0.689 & 0.032 \\
\texttt{Min-K\%++} & 0.658 & 0.028 & 0.705 & 0.030 & 0.691 & 0.034 & 0.701 & 0.034 \\
\texttt{Neighborhood} & 0.584 & 0.012 & 0.597 & 0.010 & 0.585 & 0.014 & 0.599 & 0.016 \\
\texttt{Recall} & 0.695 & 0.044 & 0.712 & 0.048 & 0.710 & \textbf{0.038} & 0.718 & 0.040 \\
\midrule
\texttt{Reference} & 0.762 & 0.012 & 0.774 & 0.010 & 0.748 & 0.006 & 0.762 & 0.008 \\
\texttt{SPV-MIA} & 0.670 & 0.042 & 0.684 & 0.046 & 0.662 & 0.028 & 0.676 & 0.032 \\
\midrule
\texttt{ICP-MIA-SP} & 0.698 & 0.056 & 0.771 & 0.086 & 0.726 & 0.026 & 0.750 & \textbf{0.054} \\
\texttt{ICP-MIA-Ref} & \textbf{0.802} & \textbf{0.058} & \textbf{0.813} & \textbf{0.092} & \textbf{0.770} & 0.028 & \textbf{0.829}& 0.050 \\
\bottomrule
\end{tabular}
\end{table}

Our analysis of the results reveals a clear relationship between the PEFT configuration and the model's susceptibility to membership inference. First, we observe that increasing the capacity of the LoRA adapter (i.e., rank) directly correlates with higher attack success rates for all methods. This provides quantitative evidence that granting the model more trainable parameters, even within a PEFT framework, increases its tendency to memorize training data, thereby enlarging the privacy attack surface. Second, quantization appears to have a mitigating effect; the QLoRA (4bit) configuration shows the lowest vulnerability, suggesting that aggressive quantization acts as a form of regularization that limits memorization.

\subsection{Impact of the Number of Context Candidates $K$}
A practical consideration for ICP-MIA is the number of probe contexts $K$ used per target sample. Because the attack selects the minimum ICP score across $K$ probes, larger $K$ can improve effectiveness but also increase API queries and computational cost. To study this trade-off, we vary $K \in \{5, 10, 15, 20, 50\}$ across all three datasets.

As shown in Appendix Figure~\ref{fig:topk_ablation}, attack performance shows clear diminishing returns as $K$ increases. For \texttt{ICP-MIA-Ref} (Figure~\ref{fig:topk_ablation_b}), accuracy improves steadily up to $K=20$, with larger values providing only marginal gains. \texttt{ICP-MIA-SP} (Figure~\ref{fig:topk_ablation_a}) exhibits the same pattern: performance increases from $K=5$ to $K=20$, after which improvements plateau. At $K=20$, ICP-MIA-SP already achieves strong results—AUC of 0.869 on CNN-DM, 0.855 on MedInstruct, and 0.836 on HealthcareMagic.
Appendix Table~\ref{tab:k_results} compares attack performance under our default settings ($K$=5 for ICP-MIA-SP and $K$=10 for ICP-MIA-Ref) against the $K$=20 setting, showing the additional gains achievable with more candidate probes. This comparison highlights the trade-off between attack performance and query cost when choosing $K$.

\subsection{Evaluating Attack Performance Under Differential Privacy}

Differential Privacy (DP)\cite{dwork2014algorithmic} provides formal privacy guarantees by injecting calibrated noise during training, thereby limiting what can be inferred about any individual training example. DP-SGD\cite{abadi2016deep}, the most widely used DP mechanism for deep learning, enforces privacy by clipping per-sample gradients and adding Gaussian noise. However, applying DP-SGD to large language models is computationally expensive. Due to resource constraints, we evaluate a smaller model—LLaMA-3.2-1B-Instruct—fine-tuned with LoRA ($r=32$) under DP-SGD. We train for 3 epochs with a learning rate of $1\text{e}{-5}$ and a cosine annealing schedule, and consider privacy budgets $\epsilon \in {10, 50, 100}$. As shown in Table~\ref{tab:dp_results}, DP-LoRA significantly suppresses all MIA methods, with AUC scores approaching random guessing (0.5) under strong privacy settings (e.g., $\epsilon=10$). This behavior is expected: both LoRA and DP-SGD reduce memorization, and their combination further weakens attack signals. Nonetheless, \texttt{ICP-MIA-SP} consistently outperforms most baselines (including all reference-free methods) across all tested privacy budgets, demonstrating its advantages even under strong DP protection.

\begin{table}[htbp]
\centering
\caption{MIA AUC with different level DP-SGD}
\label{tab:dp_results}
\begin{tabular}{lccc}
\toprule
\textbf{MIA Method} & \textbf{$\epsilon=10$} & \textbf{$\epsilon=50$} & \textbf{$\epsilon=100$} \\
\midrule
\texttt{Loss Attack} & 0.5080 & 0.5083 & 0.5121 \\
\texttt{Zlib} & 0.5012 & 0.5013 & 0.5064 \\
\texttt{Min-K\%} & 0.5184 & 0.5189 & 0.5228 \\
\texttt{Min-K\%++} & 0.5223 & 0.5225 & 0.5285 \\
\texttt{Neighborhood} & 0.5075 & 0.5080 & 0.5118 \\
\texttt{Recall} & 0.5150 & 0.5190 & 0.5232 \\
\midrule
\texttt{Reference Attack(Ref)} & 0.5160 & 0.5200 & 0.5248 \\
\texttt{SPV-MIA} & \textbf{0.5244} & 0.5302 & 0.5332 \\
\midrule
\texttt{ICP-MIA-SP} & 0.5235 & \textbf{0.5310} & \textbf{0.5367} \\
\texttt{ICP-MIA-REF} & 0.5112 & 0.5208 & 0.5244 \\
\bottomrule
\end{tabular}
\end{table}


\section{Discussion}
\label{sec:discussion}
This section discusses three additional factors affecting ICP-MIA’s effectiveness in practical deployments: (1) its applicability under label-only API constraints, (2) the impact of residual memorization when test samples come from the pre-training corpus, and (3) how training dynamics create heterogeneous vulnerability across samples.

\subsection{Label-only Attack}
\label{sec:discussion_proprietary}
While our threat model focuses on models that expose token-level log-probabilities, many commercial LLM APIs, including OpenAI (GPT series) and Anthropic (Claude), only expose label-level or text-only outputs.
Although label-only attacks are not the primary focus of this work, recent research has shown promising approaches under this setting. In particular, PETAL~\cite{he2025towards} approximates log-probabilities from token-level semantic similarity using a surrogate model and a learned regression, which enables effective membership inference without probability access. ICP-MIA naturally supports such adaptation: our method requires only a conditional scoring mechanism and does not rely on gradients or parameter updates. By replacing log-probability changes with semantic similarity–based scores, our in-context probing mechanism can operate in label-only scenarios. 

We adapted \texttt{ICP-MIA-SP} using the PETAL framework. As shown in Table~\ref{tab:label_only}, our method consistently outperforms PETAL in AUC across all datasets on LLaMA-3.2-3B-Instruct, demonstrating that ICP-MIA remains effective even when restricted to label-only access.

\begin{table}[htbp]
\centering
\caption{MIA AUC in Label-Only Setting}
\label{tab:label_only}
\begin{tabular}{lccc}
\toprule
\textbf{MIA Method} & \textbf{Healthcare} & \textbf{MedInstruct} & \textbf{CNN-DM} \\
\midrule
\texttt{PETAL} & 0.7354 & 0.7756 & 0.9018 \\
\texttt{ICP-MIA-Ref} & 0.7861 & 0.7632 & 0.7661 \\
\texttt{ICP-MIA-SP} & \textbf{0.9143} & \textbf{0.7812} & \textbf{0.9351} \\
\bottomrule
\end{tabular}
\end{table}

\subsection{The Impact of Test Samples Overlapping with the Pre-training Corpus}
\label{sec:residual_memorization}
Our threat model assumes that the fine-tuning data is private and disjoint from the model’s pre-training corpus, which is typical in many real-world deployments. However, corner cases may arise when the fine-tuning data domain partially overlaps with the pre-training data.
If a test sample appears in both datasets, the model effectively ``sees'' it twice—once during pre-training and again during fine-tuning—which can artificially increase the true positive rate. The more concerning case is when a sample appears only in the pre-training corpus, not members of our targeted fine-tuning dataset. Such residual memorization can increase the false positive rate by making pre-trained non-members look more like fine-tuning members. We focus on this latter scenario.

\begin{figure}[ht]
    \centering
    \includegraphics[width=1\linewidth]{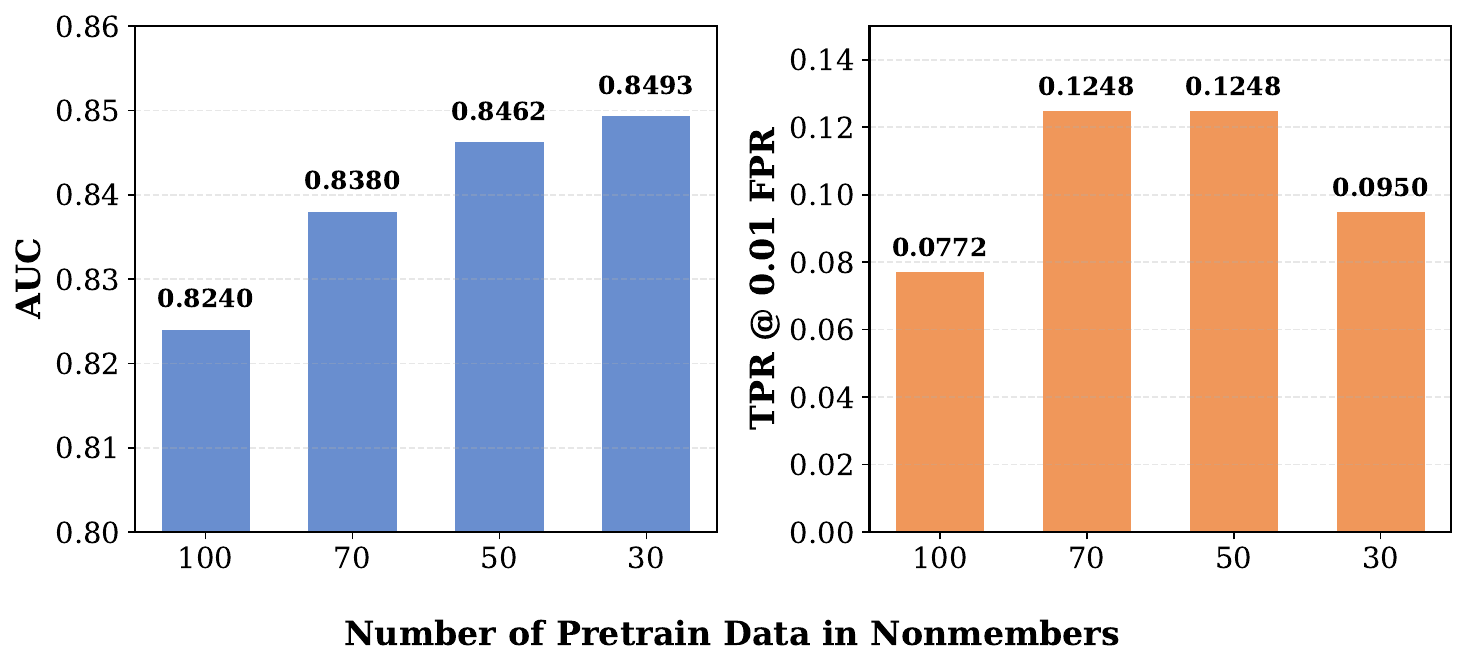}
    \caption{The Impact of Residual Memorization of Non-Members}
    \label{fig:memory_residue}
\end{figure}

We test \texttt{ICP-MIA-SP} on Pythia-2.8B-deduped with Healthcaremagic dataset by progressively replacing portions of the 500 non-member test samples with data from WikiMIA~\cite{shidetecting} (constructed from training data of Pythia models). As shown in Figure \ref{fig:memory_residue}, attack performance degrades only marginally as the proportion of pre-trained samples increases. This is consistent with our expectation, and also indicates that the effect of residual memorization is not strong enough to undermine ICP-MIA in practice.

\subsection{The Impact of Training Dynamics on MIA Vulnerability}
ICP-MIA is motivated by the Optimization Gap induced by the fine-tuning process. A natural question is therefore: how do training dynamics shape sample-level vulnerability? We investigate two key factors—training order and learning rate schedule—to understand how different configurations produce heterogeneous susceptibility to membership inference.

\noindent\textbf{Experimental Setting}
We partition the training set equally into 10 subsets, denoted $D_0, D_1, \dots, D_9$. Within each epoch, these partitions are presented to training in a fixed sequence ($D_0 \to D_9$). After each epoch, we save a model checkpoint. For each partition $D_k$, we compute an AUC score for each partition $D_k$ on a balanced test set (2,000 members from $D_k$ and 2,000 non-members). We use two different learning rate schedules: fixed and cosine annealing.

\noindent\textbf{Training under Fixed Learning Rate.}
Under a fixed learning rate, we observe a strong recency effect: vulnerability increases with the partition index in the training sequence. As shown in Appendix Figure~\ref{fig:training_dynamics_sub1}, the final partitions (e.g., $D_9$) are substantially more vulnerable than early ones, since updates to later partitions are overwritten less and therefore leave a stronger imprint on the final model.

\noindent\textbf{Training under Cosine Annealing Learning Rate.}
Using a cosine-annealing schedule yields more complex vulnerability patterns (Appendix Figure~\ref{fig:training_dynamics_sub2}).
In a single epoch, vulnerability continues to rise even after the learning rate reaches its peak, as both recency effects and still-high learning rates jointly contribute to memorization. 
As the learning rate further decreases, vulnerability drops sharply because updates to later partitions become too small to induce memorization. Extending training to two epochs produces a similar non-monotonic, single-peaked trend driven by the interplay between recency and the decaying learning rate. Additional ablations using shuffled partition orders (Appendix Figure~\ref{fig:ahsTS}) confirm that these patterns are driven by training dynamics rather than properties of specific data subsets.

\section{Conclusion}
In this paper, we propose ICP-MIA, a novel membership inference framework for fine-tuned LLMs that leverages the optimization gap between member and non-member samples. By introducing ICP as a training-free proxy for measuring residual learning potential, we designed a principled attack that operates effectively in black-box settings. Our framework supports both reference-based and reference-free probing strategies, enabling strong performance even without access to auxiliary data. Extensive experiments demonstrate that ICP-MIA achieves state-of-the-art results under strict false-positive constraints and offers actionable insights for privacy auditing in real-world LLM deployments.

\section*{Acknowledgment}
This work was supported by IBM through the IBM-Rensselaer AIRC 2024 Award.

\section{Ethics Considerations}
This study adheres to responsible disclosure and ethical AI research principles. Our membership inference attacks are conducted on publicly available datasets and models, strictly for the purpose of evaluating and improving model privacy. We do not attempt to deanonymize or identify any individuals behind the data samples. The goal is to highlight potential vulnerabilities and promote the development of privacy-preserving techniques. All experiments were performed in controlled environments, and no real user data or proprietary models were accessed without authorization.

\bibliographystyle{IEEEtran} 
\bibliography{NDSSreference}

\appendix
\subsection{Implement Details}
\label{app:Experiment_Details}
\paragraph{Fine-tuning Details.}

We performed our fine-tuning experiments using the LLaMA Factory framework. We employed a full fine-tuning paradigm with the following hyperparameters: a maximum sequence length of 2048 tokens, a learning rate of 2e-4 for Healthcaremagic; 1e-5 for MedInstruct and CNN-DM with a cosine learning rate scheduler, and a warmup ratio of 0.1. Models were trained for 2 epochs with a global batch size of 16. All training was conducted on a single NVIDIA H100 GPU. Following standard practice for SFT, the loss was calculated exclusively on the response tokens; the input prompt tokens were masked out and did not contribute to the loss.

\paragraph{Dataset Details.}
For each of the three datasets, we designated 80\% of the data for training (the member set), 10\% as the test set (the non-member set), and the remaining 10\% as a validation set. This validation set was used either as a reference pool for attacks like \texttt{ICP-MIA-Ref} or to train reference models for baseline comparisons. For each attack evaluation, we constructed a balanced test cohort by randomly sampling 1,000 data points, consisting of 500 members from the training set and 500 non-members from the test set.

\paragraph{Attack Configurations.}
Our fine-tuning follows an instruction-following format. Consequently, all attacks are evaluated based on the conditional log-likelihood of the response given the prompt. For the baseline methods, we used the following configurations:
\begin{itemize}
\item \texttt{Min-K\% and Min-K\%++}: We adopted the default setting of $k=20$ from their original papers.
\item \texttt{ReCaLL}: To ensure a fair comparison, the \texttt{ReCaLL} baseline was configured to use the same prefix pool as our \texttt{ICP-MIA-Ref} method—the Dolly-15k instruction dataset. Following their official implementation, we used 7-shot prefixes.
\item \texttt{Neighborhood Attack}: We set the hyperparameters as follows: roberta-base as masked language model, neighbors=20, and top\_k=5.

\item \texttt{Reference Attack}: We consider two configurations of the Reference Attack. Reference Attack (Base) uses the original pre-trained model as the reference, requiring no additional data. Reference Attack (Ref) constructs the reference model by fine-tuning on a held-out dataset sampled from the same distribution as the target's training data. This held-out set constitutes 10\% of the original dataset and maintains strict separation from both training and test partitions. The membership score is computed as $LL_{target}(y|x) - LL_{ref}(y|x)$. We assume that the attacker has no knowledge about the architecture of the target model. For all experiments requiring training a reference model, we use Qwen3-0.6B as the base model.

\item \texttt{SPV-MIA}: We use the same setup as the original work: T5-large for masking (span length=2, mask ratio=30\%), generating 10 perturbations per sample. The reference model is trained on the held-out set with same hyperparameters as the target model.

\item \texttt{PETAL}: We use the pretrained model of the target model as the surrogate model. We employ greedy decoding with a maximum generation length of 64 tokens and compute semantic similarity between predicted and ground-truth tokens using sentence-transformers/all-MiniLM-L6-v2.

\end{itemize}

\subsection{Experiments on Additional Datasets}
\label{app:extradataset}
We conducted additional experiments on XSum~\cite{xsum} and AG News~\cite{zhang2015characterlevel}, two public benchmarks commonly used in prior MIA evaluations~\cite{mattern2023neighborhood, fu2024practicalmembershipinferenceattacks}. We formulated XSum as a summarization task and AG News as a news completion task, performing full fine-tuning on LLaMA-3.2-3B, LLaMA-3.2-3B-Instruct, and Pythia-2.8B-deduped for 2 epochs with a learning rate of 1e-5 and cosine annealing schedule. For both ICP-MIA variants, we use $K=20$ candidate probe contexts.

As shown in Table~\ref{tab:xsum_results}, ICP-MIA-SP outperforms all existing reference-free methods and also the Reference Attack across all configurations in AUC, while achieving performance very close to SPV-MIA despite requiring no reference-model training. On XSum with LLaMA-3.2-3B-Instruct, ICP-MIA-SP attains an AUC of 0.934, close to SPV-MIA (0.945) and well above baselines such as Min-K\% (0.744) and ReCaLL (0.849). On AG News, ICP-MIA-SP reaches 0.905, slightly surpassing SPV-MIA (0.903) and substantially outperforming all reference-free baselines.

\begin{table}[ht]
\small
\centering
\caption{MIA AUC on XSum and AG News}
\label{tab:xsum_results}
\setlength{\tabcolsep}{4pt} 
\begingroup
\renewcommand{\arraystretch}{1.2}
\definecolor{lightgray}{gray}{0.9}
\resizebox{\columnwidth}{!}{%
\begin{tabular}{l|cc|cc|cc}
\toprule
\multirow{2}{*}{\textbf{MIA Method}} & \multicolumn{2}{c|}{\textbf{LLaMA-3.2-3B-Instruct}} & \multicolumn{2}{c|}{\textbf{LLaMA-3.2-3B}} & \multicolumn{2}{c}{\textbf{Pythia-2.8B-deduped}} \\
\cmidrule(lr){2-3} \cmidrule(lr){4-5} \cmidrule(lr){6-7}
 & \textbf{XSum} & \textbf{AGNews} & \textbf{XSum} & \textbf{AGNews} & \textbf{XSum} & \textbf{AGNews} \\
\midrule
\texttt{Loss Attack} & 0.765 & 0.594 & 0.800 & 0.612 & 0.780 & 0.658 \\
\texttt{Zlib} & 0.755 & 0.586 & 0.788 & 0.603 & 0.767 & 0.661 \\
\texttt{Min-K\%} & 0.744 & 0.603 & 0.799 & 0.617 & 0.767 & 0.659 \\
\texttt{Min-K\%++} & 0.694 & 0.570 & 0.730 & 0.574 & 0.703 & 0.604 \\
\texttt{Neighborhood} & 0.612 & 0.580 & 0.604 & 0.581 & 0.617 & 0.601 \\
\texttt{Recall}& 0.849 & 0.743 & 0.884 & 0.733 & 0.859 & 0.713 \\
\midrule
\texttt{Reference}& 0.883 & 0.832 & 0.865 & 0.844 & 0.831 & 0.786 \\
\texttt{SPV-MIA} & 0.945 & 0.903 & 0.920 & 0.901 & 0.892 & 0.870 \\
\midrule
\texttt{ICP-MIA-Ref} & 0.839 & 0.797 & 0.802 & 0.767 & 0.774 & 0.753 \\
\texttt{ICP-MIA-SP} & 0.934 & 0.905 & 0.941 & 0.909 & 0.885 & 0.879 \\
\bottomrule
\end{tabular}%
}
\endgroup
\end{table}

\subsection{Example of a masking-based context probe}
\label{app:example}
\small
\begin{promptbox}
\textbf{$x_{\text{prompt}}:$} Determine if the described symptoms relate to cystic fibrosis based on provided genetic information.

\textbf{$x_{\text{input}}:$} The patient exhibits regular bouts of persistent coughing, recurrent lung infections, and difficulty...

\textbf{$y:$} The described symptoms of regular bouts of persistent coughing, recurrent lung infections, and difficulty in...

\textbf{$C:$} The described symptoms of regular [MASK] of [MASK] [MASK] recurrent lung [MASK] and [MASK] in...
\end{promptbox}

\begin{table*}[ht]
\centering
\caption{Performance of ICP-MIA with default setting vs. $K=20$  }
\label{tab:k_results}
\setlength{\tabcolsep}{4pt} 
\begingroup
\renewcommand{\arraystretch}{1.2}
\definecolor{lightgray}{gray}{0.9}
\definecolor{lightblue}{RGB}{173,216,230}
\resizebox{\textwidth}{!}{%
\begin{tabular}{l|ccc|ccc|ccc}
\toprule
\multirow{3}{*}{\textbf{MIA Method}} & \multicolumn{9}{c}{\textbf{AUC}} \\
\cmidrule{2-10}
 & \multicolumn{3}{c|}{\textbf{LLaMA-3.2-3B-Instruct}} & \multicolumn{3}{c|}{\textbf{LLaMA-3.2-3B}} & \multicolumn{3}{c}{\textbf{Pythia-2.8B-deduped}} \\
\cmidrule(lr){2-4} \cmidrule(lr){5-7} \cmidrule(lr){8-10}
 & \textbf{Healthcare} & \textbf{MedInstruct} & \textbf{CNN-DM} & \textbf{Healthcare} & \textbf{MedInstruct} & \textbf{CNN-DM} & \textbf{Healthcare} & \textbf{MedInstruct} & \textbf{CNN-DM} \\
\midrule

\cellcolor{lightgray}\texttt{ICP-MIA-Ref(default)} & \cellcolor{lightgray}0.827 & \cellcolor{lightgray}0.838 & \cellcolor{lightgray}0.890 & \cellcolor{lightgray}0.842 & \cellcolor{lightgray}0.775 & \cellcolor{lightgray}0.837 & \cellcolor{lightgray}0.850 & \cellcolor{lightgray}0.746 & \cellcolor{lightgray}0.706 \\

\cellcolor{lightgray}\texttt{ICP-MIA-SP(default)} & \cellcolor{lightgray}0.942 & \cellcolor{lightgray}0.959 & \cellcolor{lightgray}0.965 & \cellcolor{lightgray}0.763 & \cellcolor{lightgray}0.977 & \cellcolor{lightgray}0.927 & \cellcolor{lightgray}0.853 & \cellcolor{lightgray}0.882 & \cellcolor{lightgray}0.845 \\

\cellcolor{lightgray}\texttt{ICP-MIA-Ref($K=20$)} & \cellcolor{lightgray}0.821 & \cellcolor{lightgray}0.855 & \cellcolor{lightgray}0.869 & \cellcolor{lightgray}0.845 & \cellcolor{lightgray}0.789 & \cellcolor{lightgray}0.834 & \cellcolor{lightgray}0.836 & \cellcolor{lightgray}0.751 & \cellcolor{lightgray}0.741 \\

\cellcolor{lightgray}\texttt{ICP-MIA-SP($K=20$)} & \cellcolor{lightgray}0.948 & \cellcolor{lightgray}0.962 & \cellcolor{lightgray}0.968 & \cellcolor{lightgray}0.796 & \cellcolor{lightgray}0.978 & \cellcolor{lightgray}0.936 & \cellcolor{lightgray}0.871 & \cellcolor{lightgray}0.880 & \cellcolor{lightgray}0.852 \\
\bottomrule
\end{tabular}%
}
\resizebox{\textwidth}{!}{%
\begin{tabular}{l|ccc|ccc|ccc}
\toprule
\multirow{3}{*}{\textbf{MIA Method}} & \multicolumn{9}{c}{\textbf{TPR@1\%FPR}} \\
\cmidrule{2-10}
 & \multicolumn{3}{c|}{\textbf{LLaMA-3.2-3B-Instruct}} & \multicolumn{3}{c|}{\textbf{LLaMA-3.2-3B}} & \multicolumn{3}{c}{\textbf{Pythia-2.8B-deduped}} \\
\cmidrule(lr){2-4} \cmidrule(lr){5-7} \cmidrule(lr){8-10}
 & \textbf{Healthcare} & \textbf{MedInstruct} & \textbf{CNN-DM} & \textbf{Healthcare} & \textbf{MedInstruct} & \textbf{CNN-DM} & \textbf{Healthcare} & \textbf{MedInstruct} & \textbf{CNN-DM} \\
\midrule

\cellcolor{lightgray}\texttt{ICP-MIA-Ref(default)} & \cellcolor{lightgray}0.084& \cellcolor{lightgray}0.044 & \cellcolor{lightgray}0.020 & \cellcolor{lightgray}0.140 & \cellcolor{lightgray}0.018 & \cellcolor{lightgray}0.062 & \cellcolor{lightgray}0.110 & \cellcolor{lightgray}0.074 & \cellcolor{lightgray}0.022 \\

\cellcolor{lightgray}\texttt{ICP-MIA-SP(default)} & \cellcolor{lightgray}0.172 & \cellcolor{lightgray}0.326 & \cellcolor{lightgray}0.518 & \cellcolor{lightgray}0.070 & \cellcolor{lightgray}0.538 & \cellcolor{lightgray}0.418 & \cellcolor{lightgray}0.122 & \cellcolor{lightgray}0.270 & \cellcolor{lightgray}0.144 \\

\cellcolor{lightgray}\texttt{ICP-MIA-Ref($K=20$)} & \cellcolor{lightgray}0.114 & \cellcolor{lightgray}0.116 & \cellcolor{lightgray}0.018 & \cellcolor{lightgray}0.146 & \cellcolor{lightgray}0.022 & \cellcolor{lightgray}0.008 & \cellcolor{lightgray}0.124 & \cellcolor{lightgray}0.152 & \cellcolor{lightgray}0.044 \\
\cellcolor{lightgray}\texttt{ICP-MIA-SP($K=20$)} & \cellcolor{lightgray}0.215 & \cellcolor{lightgray}0.410 & \cellcolor{lightgray}0.696 & \cellcolor{lightgray}0.069 & \cellcolor{lightgray}0.604 & \cellcolor{lightgray}0.374 & \cellcolor{lightgray}0.200 & \cellcolor{lightgray}0.324 & \cellcolor{lightgray}0.140 \\
\midrule
\bottomrule
\end{tabular}%
}
\endgroup
\vspace{-10pt}
\end{table*}

\subsection{Context Probe Example for ICP-MIA-REF}
\label{app:probe-ref-example}
\begin{promptbox}
  \textbf{Instruction:} If you are a doctor, please answer the medical questions based on the patient's description.\\
  \textbf{Question:} I woke up this morning feeling the whole room is spinning when i was sitting down. I went to the bathroom walking unsteadily, as i tried to focus i feel nauseous. I try to vomit but it wont come out.. After taking panadol and sleep for few hours, i still feel the same....\\
  \textbf{Answer:} Hi, Thank you for posting your query. The most likely cause for your symptoms is benign paroxysmal positional vertigo (BPPV), a type of peripheral vertigo. In this condition, the most common symptom is dizziness or giddiness, which is made worse with movements. ...\\
  
  \textbf{In-Context Probe:} \\
    \textbf{Instruction}: "If you are a doctor, please answer the medical questions based on the patient's description."\\
    \textbf{Question}: "Hello doctor, After unsafe exposure, I got 49 days ELISA antibody test done, 71 days HIV proviral DNA PCR test, 87 days ELISA antibody test. All were negative. The antibody test I took is not a fourth generation test. Is it conclusive or should I take another test?...."\\
    \textbf{Answer}: "Hi. Your tests are conclusive and you are not infected. No need for further tests. Ear ache and tongue papilla are not due to HIV and may be a simple bacterial infection...."
\end{promptbox}

\subsection{Prompt for perturbation generation}
\label{app:prompt_perturbation_generation}
\begin{promptbox}
\textbf{System:}"You are a precise editor. Given the original text, generate a new text in which exactly 20 words are changed (added, removed, or replaced), but the overall meaning remains identical. Do not change more than 20 tokens. Output only the new text."
\textbf{User:} ``Original text:''
\end{promptbox}

\begin{figure}[ht]
    \centering
    \includegraphics[width=1\linewidth]{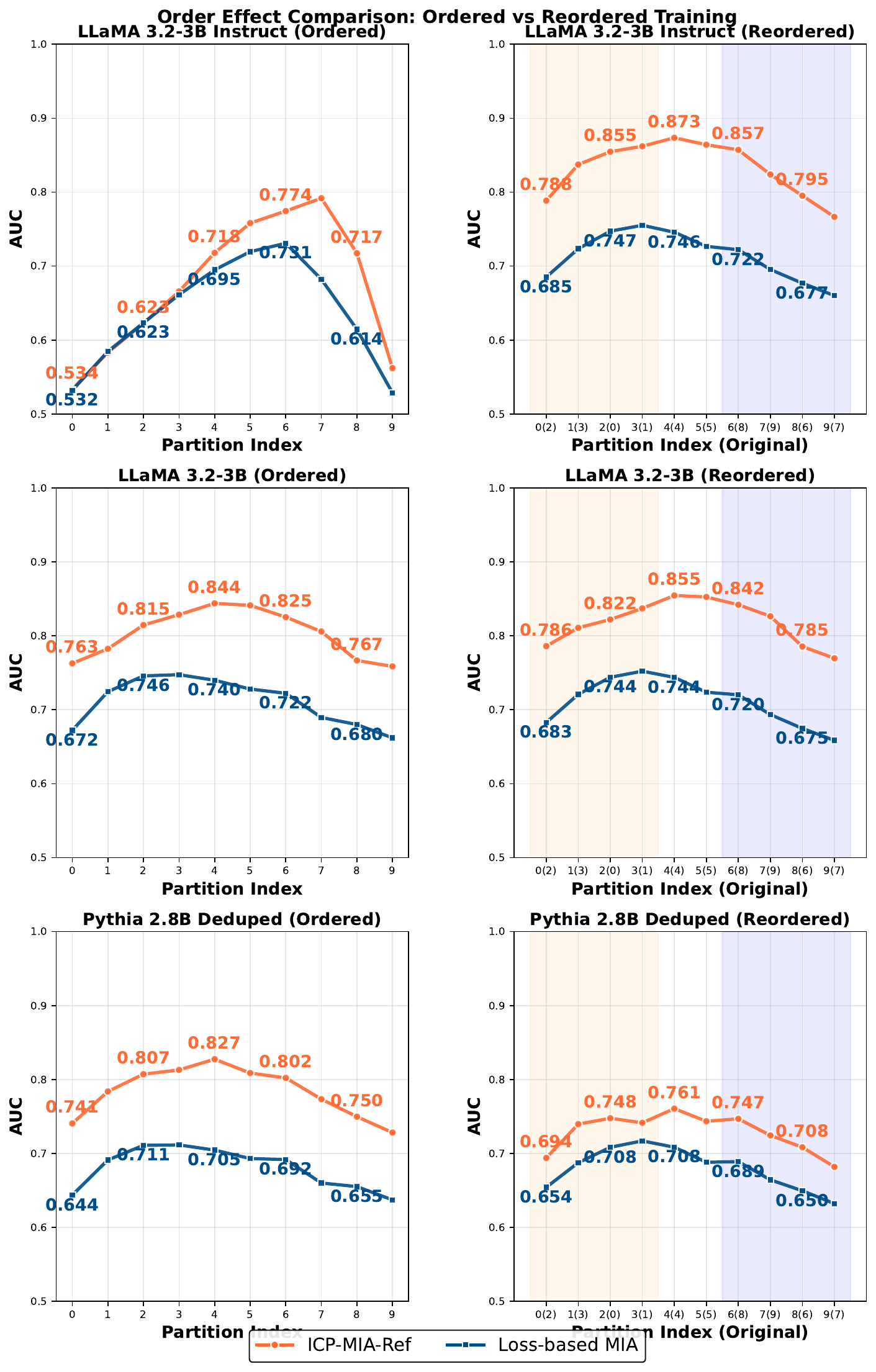}
    \caption{The Impact of Training Sequence}
    \label{fig:ahsTS}
        \vspace{-10pt}
\end{figure}

\begin{figure*}[ht]
    \centering
    \begin{subfigure}[b]{0.8\linewidth}
        \centering
        \includegraphics[width=\linewidth]{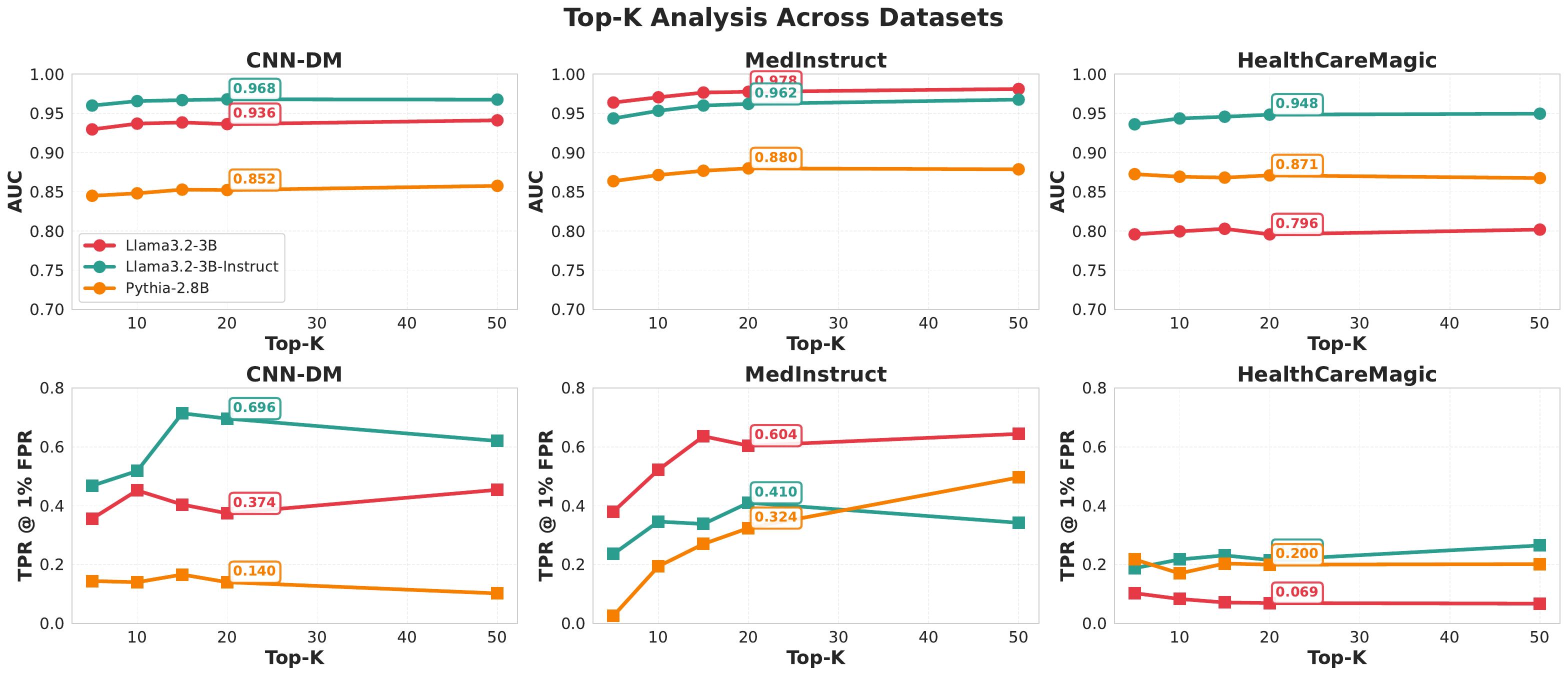}
        \caption{ICP-MIA-SP}
        \label{fig:topk_ablation_a}
    \end{subfigure}
    \begin{subfigure}[b]{0.8\linewidth}
        \centering
        \includegraphics[width=\linewidth]{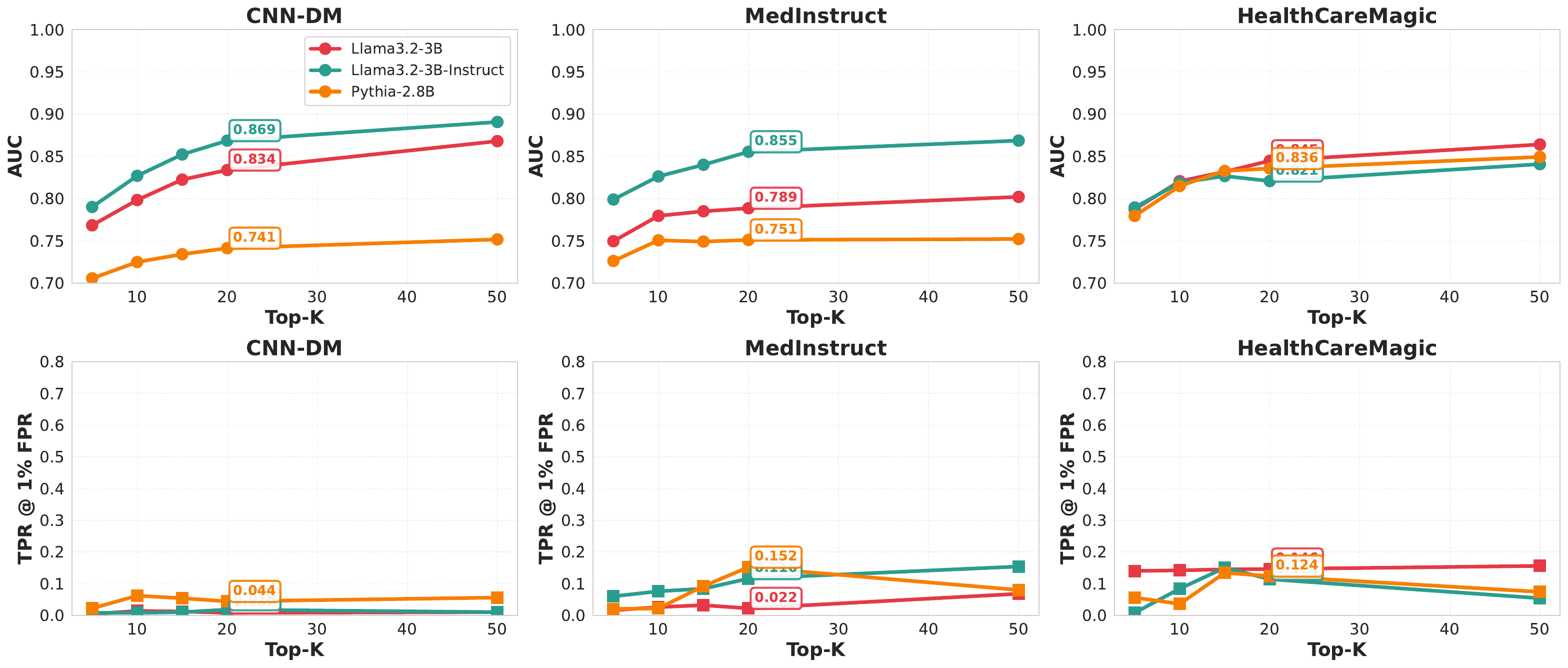}
        \caption{ICP-MIA-Ref}
        \label{fig:topk_ablation_b}
    \end{subfigure}
    \caption{Efficiency analysis on Top K}
    \label{fig:topk_ablation}
\end{figure*}

\begin{figure*}[!ht]
    \centering
    \begin{subfigure}{0.8\textwidth}
        \centering
        \includegraphics[width=\textwidth]{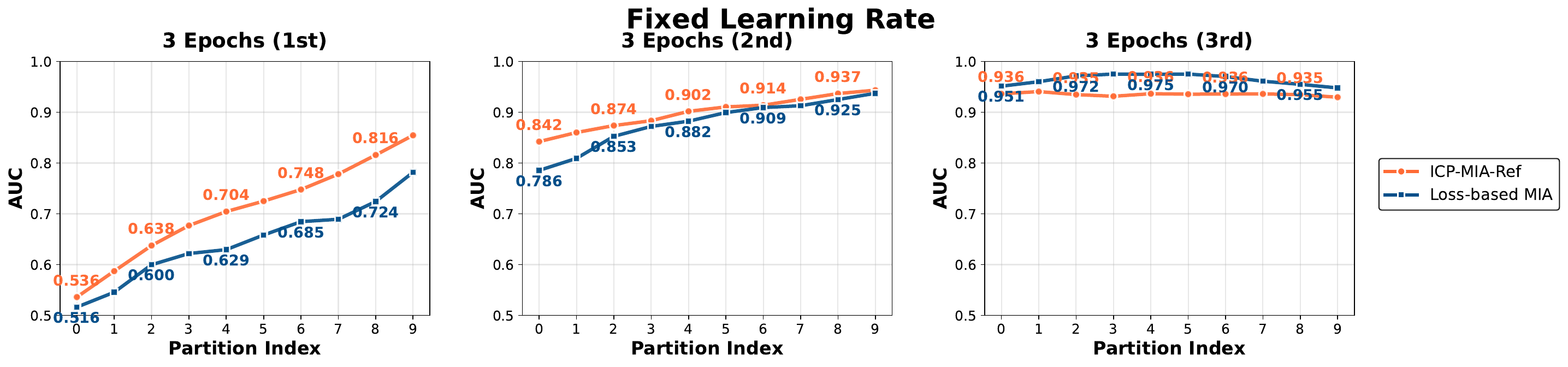}
        \caption{}
        \label{fig:training_dynamics_sub1}
    \end{subfigure}
    \begin{subfigure}{0.8\textwidth}
        \centering
        \includegraphics[width=\textwidth]{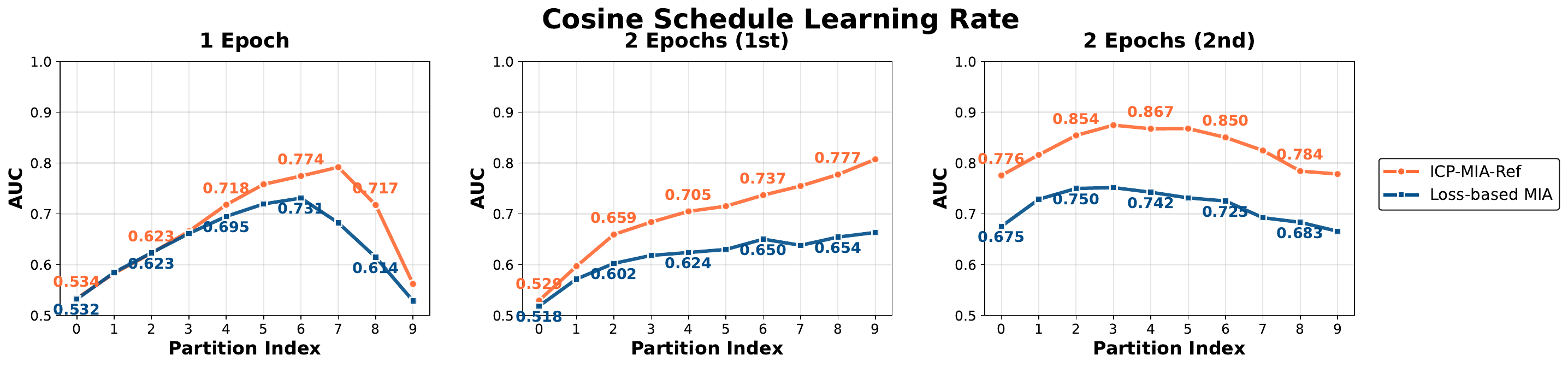}
        \caption{}
        \label{fig:training_dynamics_sub2}
    \end{subfigure}
    \caption{Training Sequence and Learning Rate Schedule Lead to Different MIA Vulnerability}
    \label{fig:training_dynamics}
\end{figure*}

\begin{figure*}[ht]
    \centering
    \includegraphics[width=0.85\linewidth]{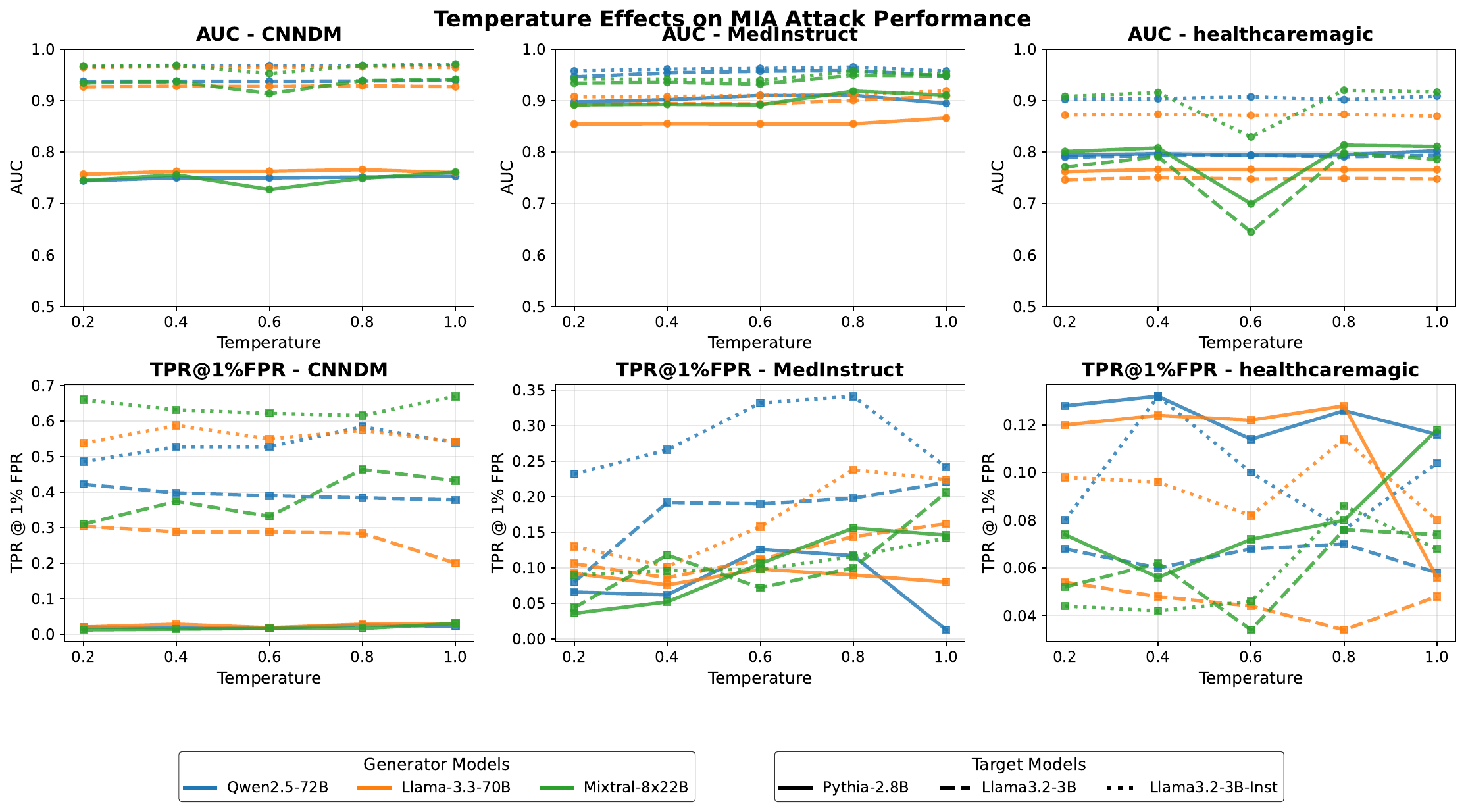}
    \caption{Ablation study on temperature in Prefix Generation}
    \label{fig:temperature_ablation}
\end{figure*}

\clearpage




\appendices
\section{Artifact Appendix}

This appendix provides instructions for reproducing the main experimental results of our paper ``In-Context Probing for Membership Inference in Fine-Tuned Language Models.'' Complete documentation is available in the \texttt{README.md} file in the repository.

\subsection{Description and Requirements}

\subsubsection{How to Access}
\url{https://doi.org/10.5281/zenodo.17906756}

\subsubsection{Hardware Dependencies}
\begin{itemize}[nosep,leftmargin=*]
    \item \textbf{Minimal:} NVIDIA A100 (80\,GB), 64\,GB RAM, 140\,GB storage.
    \item \textbf{Recommended:} NVIDIA H100 (80\,GB), 128\,GB RAM, 200\,GB storage (used in paper). Batch size is adjustable for different GPU memory.
\end{itemize}

\subsubsection{Software Dependencies}
\begin{itemize}[nosep,leftmargin=*]
    \item \textbf{Core:} Python 3.10+, CUDA 12.1+
    \item \textbf{Key packages:} PyTorch 2.5.1, Transformers 4.57.0, Datasets 4.1.1, FAISS-CPU 1.9.0, Sentence-Transformers 5.1.1.
\end{itemize}
\noindent Complete list in \texttt{requirements.txt}. LLaMA-Factory is used for fine-tuning.

\subsubsection{Benchmarks}
\begin{itemize}[nosep,leftmargin=*]
    \item \textbf{Primary dataset:} HealthCareMagic-100k.
    \item \textbf{Reference dataset:} Dolly-15k.
    \item \textbf{Model:} LLaMA-3.2-3B-Instruct (requires HuggingFace account and license).
    \item \textbf{Optional:} CNN-DM, iCliniq, MedInstruct-52k, TOFU.
\end{itemize}

\subsection{Installation and Configuration}

Detailed installation instructions in \texttt{README.md}. Summary:
\begin{enumerate}[nosep,leftmargin=*]
    \item Create conda environment with Python 3.10.
    \item Install dependencies: \texttt{pip install -r requirements.txt}
    \item Install LLaMA-Factory (see README).
    \item Login to HuggingFace: \texttt{huggingface-cli login}
\end{enumerate}

\subsection{Experiment Workflow}

The workflow consists of four stages:
\begin{enumerate}[nosep,leftmargin=*]
    \item Data preparation
    \item Model fine-tuning (3 epochs; use 2nd epoch checkpoint)
    \item Perturbation generation
    \item Attack execution
\end{enumerate}
Each stage has a corresponding script with configuration files. Fine-tuning is performed in a \texttt{LLamaFactory} environment, while attacks are executed in a separate \texttt{ICPMIA} environment.

\subsection{Major Claims}

\begin{enumerate}[label=\textbf{C\arabic*:},leftmargin=*]
    \item ICP-MIA-SP significantly outperforms baselines (AUC: 0.942 vs ReCaLL: 0.847, Min-K\%: 0.837). Validated by E1, Table~1.
    \item High-precision performance (TPR@1\%FPR: 0.172 and 0.084). Validated by E1, Table~1.
    \item ICP-MIA-SP is practical and reference-free. Validated by E1, Table~1.
    \item ICP-MIA-Ref achieves competitive performance with general-purpose reference data (AUC: 0.827). Validated by E1, Table~1.
\end{enumerate}

\subsection{Evaluation}

\subsubsection{Experiment E1: Main Results on HealthCareMagic}

\noindent\textbf{Time:} 15 human-minutes + 8--9 compute-hours

\medskip\noindent
This experiment reproduces Table~1 results (HealthCareMagic column, LLaMA-3.2-3B-Instruct) and validates claims C1--C4.

\paragraph{Preparation}

\textbf{Step 1} [5\,min]: Download and split data.
\begin{lstlisting}
python prepare_data.py \
  --dataset lavita/ChatDoctor-HealthCareMagic-100k \
  --output_dir ./data/healthcaremagic
\end{lstlisting}
Creates train/val/test splits in \texttt{./data/healthcaremagic/}.

\medskip
\textbf{Step 2} [2\,min]: Copy data files to LLaMA-Factory and add dataset entries to \texttt{dataset\_info.json}. See \texttt{README.md} for details.

\noindent For HealthCareMagic:
\begin{lstlisting}
"healthcaremagic_train": {"file_name": "healthcaremagic_train.json"},
"healthcaremagic_val":   {"file_name": "healthcaremagic_val.json"},
"healthcaremagic_test":  {"file_name": "healthcaremagic_test.json"}
\end{lstlisting}

\medskip
\textbf{Step 3} [$\sim$8\,hours]: Fine-tune model for 3 epochs using LLaMA-Factory.
\begin{lstlisting}
cd LLaMA-Factory
llamafactory-cli train ../config/config_training_Healthcare.yaml
\end{lstlisting}
The model checkpoints are saved in the \texttt{saves/} directory. \textbf{Use the 2nd epoch checkpoint} (e.g., \texttt{checkpoint-XXX}) as the target model for the attack.

\medskip
\textbf{Step 4} [5\,min]: Generate perturbations for the attack dataset.
\begin{lstlisting}
python generate_perturbations.py convert \
  --input ./data/healthcaremagic/healthcaremagic_attack.json \
  --output ./data/healthcaremagic/healthcaremagic_attack_perturbed.json \
  --mask_rate 0.7 --num_perturbations 20
\end{lstlisting}

\paragraph{Execution}

\textbf{Step 5} [$\sim$20\,min]: Update \texttt{target\_model\_path} in config files to point to the 2nd epoch checkpoint, then run attacks.
\begin{lstlisting}
# Self-perturbation ICP-MIA
python icp_mia_attack.py \
  --config config/config_icp_sp_healthcare.yaml

# Similarity-based ICP-MIA
python icp_mia_attack.py \
  --config config/config_icp_ref_Healthcare.yaml
\end{lstlisting}

\noindent For MedInstruct, use \texttt{config\_icp\_sp\_MedInstruct.yaml} and \texttt{config\_icp\_ref\_MedInstruct.yaml} instead.

\paragraph{Results}

Results are saved in \texttt{./results/}:
\begin{itemize}[nosep,leftmargin=*]
    \item \texttt{\{exp\_name\}\_results.csv}: summary metrics (AUC, TPR@FPR).
    \item \texttt{\{exp\_name\}\_\{attack\}\_detailed\_scores.json}: raw scores.
\end{itemize}
Compare AUC and TPR@1\%FPR with Table~1 to validate claims C1--C4.

\subsubsection{Optional: Baseline Comparison}

\noindent\textbf{Time:} 30\,min

\medskip\noindent
Run baseline attacks to reproduce baseline columns in Table~1:
\begin{lstlisting}
python baseline/main.py \
  -c baseline/config/config_healthcaremagic.yaml \
  --attacks loss minkprob minkplusplus zlib reference \
  --output baseline_results.pkl
\end{lstlisting}

\subsubsection{Optional: Extended Evaluation}

The artifact supports additional experiments (other datasets, ablation studies). Modify configuration files as described in \texttt{README.md}. These are optional and not required for validating core claims.

\subsection{Notes}

\begin{itemize}[nosep,leftmargin=*]
    \item \textbf{Time:} $\sim$8--9 hours on H100 (mostly fine-tuning).
    \item \textbf{Disk space:} $\sim$140\,GB.
    \item \textbf{Reproducibility:} All experiments use \texttt{random\_seed:\ 42}. Minor variations may occur due to hardware differences.
    \item \textbf{Troubleshooting:} See \texttt{README.md} for common issues.
\end{itemize}


\end{document}